\documentclass[12pt,preprint]{aastex}

\newcommand{\Msun}{M_{\sun}}
\newcommand{\Msunyr}{\Msun\,{\mathrm{yr}}^{-1}}
\newcommand{\cm}{{\mathrm{cm}}}
\newcommand{\km}{{\mathrm{km}}}
\newcommand{\erg}{{\mathrm{erg}}}
\newcommand{\second}{{\mathrm{s}}}
\newcommand{\ergs}{{\mathrm{erg}}\,\second^{-1}}
\newcommand{\cms}{\cm\,\second^{-1}}

\newcommand{\gcc}{\mathrm{g}\,\cm^{-3}}

\slugcomment{Draft for ApJ, \today}

\shorttitle{RELATIVISTIC JETS IN COLLAPSARS}
\shorttitle{Zhang, Woosley, \& MacFadyen}

\begin{document}

\title{RELATIVISTIC JETS IN COLLAPSARS}
\author{Weiqun Zhang$^1$, S. E. Woosley$^1$, and A. I. MacFadyen$^2$}
\affil{
$^1$Department of Astronomy and Astrophysics, 
    University of California,
    Santa Cruz, CA 95064 \\
$^2$Caltech MC 130-33, 
    Pasadena, CA 91125
}

\begin{abstract}
We examine the propagation of two-dimensional relativistic jets
through the stellar progenitor in the collapsar model for gamma-ray
bursts (GRBs). Each jet is parameterized by a radius where it is
introduced, and by its initial Lorentz factor, opening angle, power,
and internal energy. In agreement with previous studies, we find that
relativistic jets are collimated by their passage through the stellar
mantle. Starting with an initial half-angle of up to 20 degrees, they
emerge with half-angles that, though variable with time, are around 5
degrees. Interaction of these jets with the star and their own cocoons
also causes mixing that sporadically decelerates the flow.  We
speculate that this mixing instability is chiefly responsible for the
variable Lorentz factor needed in the internal shock model and for the
complex light curves seen in many gamma-ray bursts. In all cases
studied, the jet is shocked deep inside the star following a brief
period of adiabatic expansion. This shock converts most of the jet's
kinetic energy into internal energy so that even initially
``cold'' jets become hot after going a short distance.  The jet that
finally emerges from the star thus has a moderate Lorentz factor,
modulated by mixing, and a very large internal energy.  In a second
series of calculations, we follow the escape of that sort of
jet. Conversion of the remaining internal energy gives terminal
Lorentz factors along the axis of approximately 150 for the initial
conditions chosen. Because of the large ratio of internal to kinetic
energy in both the jet ($\geq 80\%$) and its cocoon, the opening angle
of the final jet is significantly greater than at breakout. A small
amount of material emerges at large angles, but with a Lorentz factor
still sufficiently large to make a weak GRB. This leads us to propose
a ``unified model'' in which a variety of high energy transients,
ranging from x-ray flashes to ``classic'' GRBs, may be seen depending
upon the angle at which a standard collapsar is observed.  We also
speculate that the breakout of a relativistic jet and its collision
with the stellar wind will produce a brief transient with properties
similar to the class of ``short-hard'' GRBs.  Implications of our
calculations for GRB light curves, the luminosity-variability
relation, and the GRB-supernova association are also discussed.

\end{abstract}


\section{INTRODUCTION}

Growing evidence connects GRBs to the death of massive stars.
Analysis by \citet{fra01} and others of the radio afterglows of
``long-soft'' GRBs suggests that, despite great diversity in apparent
brightnesses, these events have a total kinetic energy in relativistic
matter tightly clustered around $3 \times 10^{51}\,\erg$ \citep{fra01}
- a supernova-like energy.  ``Bumps'' resembling the light curves of
Type I supernovae have also been seen in the optical afterglows of at
least four GRBs (GRB 980326, Bloom et al. 1999; GRB 970228, Reichart
1999, Galama et al. 2000; GRB 011121, Bloom et al. 2002, Garnavich et
al. 2002; and GRB 020405, Price et al. 2002) and GRB 980425 has been
associated with an optical supernova, SN 1998bw (e.g., Galama et
al. 1998; Iwamoto et al. 1998; Woosley, Eastman, \& Schmidt 1999). The
observational evidence that (long-soft) GRBs are associated with
regions of star formation has become overwhelming
\citep{blo02b}. Given the evidence for beaming and relativistic
motion, it is probable that at least one major subclass of GRBs is a
consequence of massive stars that, in their explosive deaths, produce
relativistic jets.

Here we examine the passage of relativistic jets through a collapsing
massive star and their breakout. We begin with a rotating star that
has already experienced 10 seconds of collapse. Ten seconds is the
nominal time for a disk to form around a black hole and the polar
region to become sufficiently evacuated for a polar jet to propagate
\citep{mac99}. The initial progenitor is a helium core of $15\,\Msun$
evolved by \citet{heg02} to iron core collapse with approximations to
all (non-magnetic) forms of angular momentum transport included. The
missing 10 seconds is followed using a non-relativistic
two-dimensional code as in \citet{mac99}.  We pick up the calculation
when the jet, which presumably began in a region $\sim 30\,\km$ in
size, has already reached a radius of $2000\,\km$ and do not consider
what has gone on inside. For present purposes, details of whether the
jet was made by black hole angular momentum, MHD processes in the
disk, or neutrinos do not concern us. The jet is initiated in a
parametric way based upon its power, opening angle, Lorentz factor,
and internal energy. Its propagation to the stellar surface at
$800,000\,\km$, and its interaction with the stellar mantle is then
followed (\S~\ref{jin}). Additional calculations (\S~\ref{jout}) also
examine what happens to the jet immediately after it escapes the star
and converts its residual internal energy into additional relativistic
motion.

We find, in agreement with \citet{alo00}, that the passage of the jet
through the star leads to its additional collimation.  We also find
that instabilities along the beam's surface lead to mixing with the
nearly stationary stellar material and cocoon. The mixing produces
variations in the mass loading, and therefore the Lorentz factor of
the jet. The opening angle of the jet also varies with time,
gradually, if irregularly, growing as the star is blown aside. These
results, discussed in \S~\ref{obs}, have important implications for
the observed light curves and energies of GRBs and imply that what is
seen may vary greatly with viewing angle. In particular, we predict
the existence of a large number of low energy GRBs with mild Lorentz
factors (\S~\ref{unified}) that may be related to GRB 980425/SN 1998bw
and to the recently discovered ``hard x-ray flashes'' \citep{hei01}.

Finally, we consider the breakout of the jet. As the shock breaks out
a small amount of material being pushed ahead by the jet head is
accelerated to relativistic speeds. Interaction of this material with
the stellar wind of the progenitor will produce a transient of some
sort \citep{woo99a, tan01}.  We speculate that this is the origin of a
hard precursor to GRBs that, at least in some cases, might be
characterized as a ``short-hard GRB'' in isolation(\S~\ref{shb}).

\section{COMPUTATIONAL PROCEDURE AND ASSUMPTIONS}

\subsection{The Relativistic Code Employed}
\label{relcode}

In order to study relativistic jets in collapsars, we developed a
special relativistic, multiple-dimensional hydrodynamics code similar
to the GENESIS code \citep{alo99}. Our program employs an explicit
Eulerian Godunov-type method with numerical flux calculated using an
approximate Riemann solver: Marquina's flux formula. The PPM algorithm
is used to reconstruct the variables so that high-order spatial
accuracy is achieved.  Second- or third-order Runga Kutta methods are
employed for the time integration.  This method has previously been
used successfully to study relativistic jets in collapsars and many
other ultra-relativistic flows \citep{alo99}.  The code has been
extensively tested and the results are comparable with those found in
the literature \citep[and references therein]{mar99}.  To model
collapsars more accurately, the code was revised to include a
realistic equation of state \citep{bli96}, that, in particular, allows
a Fermi gas of electrons and positrons with arbitrary relativity and
degeneracy. Radiation and an ideal gas of nuclei are also included.
The code was also modified to include the advection of nuclear
species.  Nuclear burning, which is not important in the present
problem, is not considered.  Because we are interested in phenomena
far from the central black hole, gravity is treated in the Newtonian
approximation. We do not include the self gravitational force of the
mass on our grid, but only a radial force from the mass inside our
inner boundary condition.

\subsection{The Stellar Model}\label{init}

Our initial model is derived from a 15 $\Msun$ helium star of 0.1
solar metallicity with an initial surface rotation rate of 10\%
Keplerian on the helium burning main sequence. This Wolf-Rayet star
might be the result of a star with initially 40 $\Msun$ on the main
sequence.  The rotating helium star is evolved to iron core collapse
(defined as when any part of the star is collapsing at $10^8\,\cms$)
including angular momentum transport, but neglecting mass loss and
magnetic fields (Heger, Langer, \& Woosley 2001; Heger \& Woosley
2002). It is presumed that this helium star has lost its hydrogen
envelope either to a companion star or by stellar winds. The
collapsing, but still initially spherical star is mapped into a
two-dimensional non-relativistic code \citep{mac99} and the inner 2
$\Msun$ of iron core, which is assumed to collapse to a black hole,
replaced by an absorbing inner boundary condition at $10^{7}\,\cm$.
Poisson's equation is then solved using a post-Newtonian gravitational
potential.  Neutrino cooling by thermal processes and pair capture are
included, as well as the photodisintegration of nuclei into nucleons
in the inner part of the disk, once one has formed.  The calculation
is run until a low density channel ($\rho \lesssim 10^6\,\gcc$) has
cleared along the rotational axis. This takes approximately
$10\,\second$ after the initial core collapse.  The collapsing rotating
star is then mapped into our two-dimensional special relativistic
code(\S~\ref{relcode}), the inner boundary moved out to $2 \times
10^{8}\,\cm$, and a $5.0\,\Msun$ point mass placed at its center.

During this $10\,\second$, rotation and infall appreciably modify the
core structure, but only inside of $5000\,\km$. Outside $10,000\,\km$,
one might just as well use a non-rotating presupernova model. The
present study with an inner boundary at $2000\,\km$ is thus moderately
sensitive to the post-collapse evolution of the rotating
progenitor. For example, the density in the equatorial plane at
$2000\,\km$ and $10\,\second$ is ten times greater than that at the
same radius along the axis.  However, this non-spherical density
structure does not critically influence jet propagation.  Our
two-dimensional starting model is available to others and might be the
basis for comparison calculations that are more sensitive to the
structure of the inner star.

\subsection{Initial Conditions}\label{initial}

A two-dimensional spherical grid ($r, \theta$) is employed consisting
of 480 logarithmically spaced radial zones, 120 uniform angular zones
in the polar region ($0\degr \leq \theta \leq 30\degr$) and 80
logarithmically spaced angular zones in the range $30\degr \leq \theta
\leq 90\degr$.  The initial model is remapped onto this grid with an
inner boundary at $2 \times 10^8\,\cm$, and an outer boundary at $9
\times 10^{10}\,\cm$.

As the initial condition for the calculation, a highly relativistic
jet is assumed to have formed interior to the inner boundary of the
computational grid.  The jet is implemented in the calculation as an
inner boundary condition. An axisymmetric outflow is injected in a
purely radial direction through the inner boundary at $2 \times
10^{8}\,\cm$ within a half-angle $\theta \leq \theta_{0}$, where
$\theta_{0}$ is a free parameter.  The processes producing the jet are
uncertain, but are presumed to have endowed it with a large energy per
baryon. At the origin this may have been in the form of a large
internal energy (as in the neutrino-version of the collapsar model or
a jet energized by magnetic reconnection) or the jet may have been
born with low entropy and highly directed motion (as in some versions
of MHD accelerated jets). As we shall see, processes deep inside the
star tend to erase the distinction.

The jet is specified by its power, $\dot{E}$, initial opening
half-angle, $\theta_{0}$, initial Lorentz factor, $\Gamma_{0}$, and
the ratio of its kinetic energy\footnote{The kinetic energy density in
the lab frame is defined as $\rho \Gamma (\Gamma - 1)$, here $\rho$ is
the local rest mass density, and $\Gamma$ is the Lorentz factor in the
lab frame} to its total energy, $f_{0}$.  Of these four parameters,
the power is the easiest to address. First, observationally, we know
that the total energy {\sl in relativistic ejecta} is of order a few
times $10^{51}\,\erg$ for the entire event (e.g., Frail et al. 2001).
The energy in subrelativistic ejecta, i.e., the supernova, is probably
comparable or greater (Woosley et al. 1999; Iwamoto et al. 1998). A
total energy of order $10^{52}\,\erg$ is thus reasonable. From another
perspective, roughly one to several solar masses of material will
accrete into the black hole during the principal phase of the GRB. If
a few tenths of a percent of the rest mass energy is used to power
jet-like outflows, one again obtains $\sim 10^{52}\,\erg$. In the lab
frame, the duration of a (long-soft) GRB is $\sim 10\,\second$, so a
jet power of $10^{51}\,\ergs$ is reasonable and probably accurate to
an order of magnitude.  We employed this energy for each jet so that
the total is $2 \times 10^{51}\,\ergs$ (or $2 \times 10^{52}\,\erg$ in
$10\,\second$).  In light of recent developments that have tended to
downgrade the total energy of a GRB to a few times $10^{51}\,\erg$
\citep{fra01} and even suggest that the kinetic energy after the GRB
is over is only a few times $10^{50}\,\erg$ \citep{pan01}, this may be
too large. So we also ran one set of calculations with an energy 30\%
as great. This is still large compared with the estimate by
\citet{pan01}.

The ratio of kinetic energy to total energy, $f_0$, depends on how the
jet was born and its history before it entered the computational grid
and is more difficult to estimate.  We assume, at $2000\,\km$, that
our jet has comparable internal and kinetic energies in the laboratory
frame. Specifically, we adopt an initial Lorentz factor, $\Gamma$, of
50 for Models JA and JB, and $f_0 = 1/3$.  These values are reasonable
for a jet which began at say $30\,\km$ consisting initially of a small
amount of matter at rest with about $\sim 150$ times its rest mass in
internal energy. That is, in the notation of \citet{mes93}, $\eta =
150$. For a jet of constant opening angle, $\theta_0$, the Lorentz
factor of an adiabatically expanding jet scales as $\sim r$.  If
$\theta_0$ decreases with $r$ inside of $2000\,\km$ as is likely,
$\Gamma$ grows more gradually.  For the parameters chosen in
Table~\ref{tab:j}, the terminal Lorentz factors of the jets if they
expanded freely to infinity would be 150 for Models JA and JB. We also
explored, for comparison, Model JC which had a lower initial Lorentz
factor and a smaller total power, but a larger fraction of internal
energy (Table~\ref{tab:j}).  Fortunately, as we shall see, the initial
partition of energy is not an important parameter of the model - so
long as the total energy per baryon is large.

\section{RESULTS}

Because of the difficulty of carrying too large a range of radii on
the computational grid, each calculation was carried out in two
stages: 1) propagation inside the star, and 2) propagation in the
stellar wind after breakout. The calculations, although
two-dimensional, did not include the dynamical effects of rotation in
the jet.

\subsection{Jet Propagation Inside the Star}\label{jin}

\begin{figure}
\epsscale{1.0}
\plotone{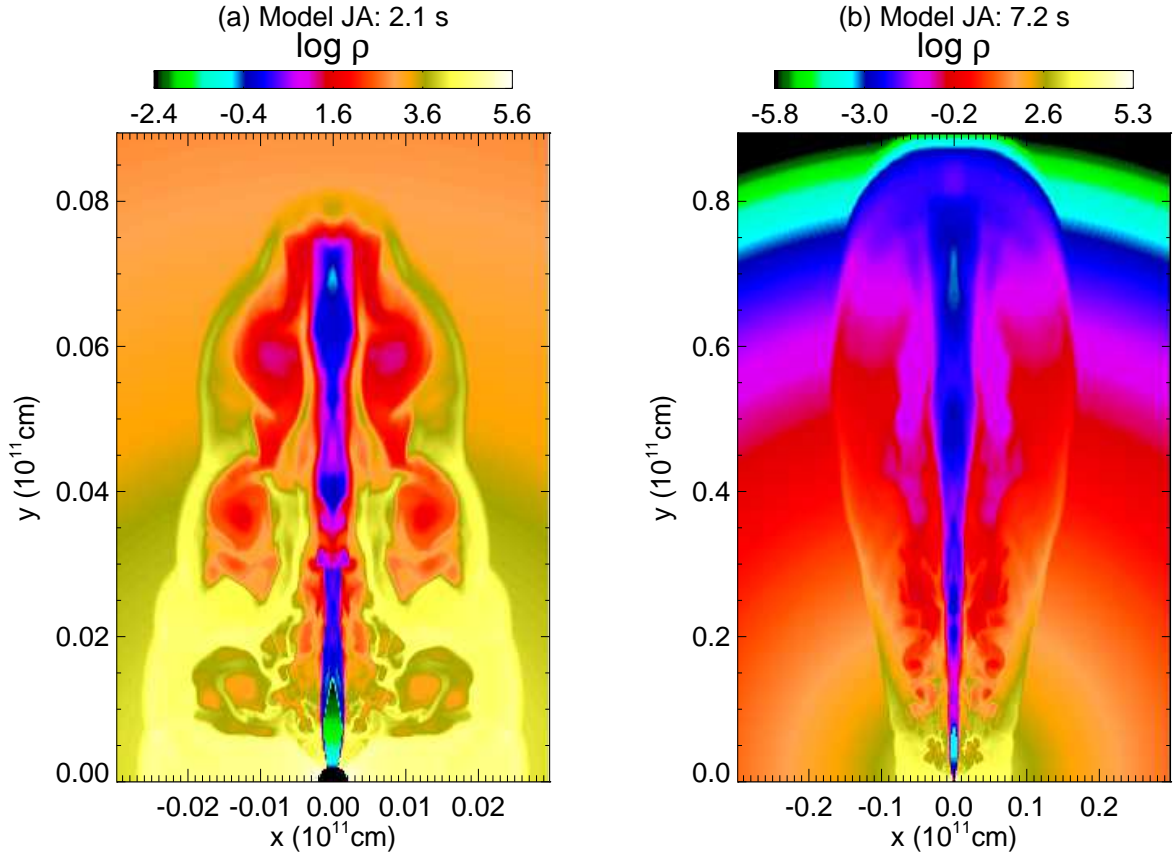}
\caption{Density structure in the local rest frame for Model JA at (a)
$t = 2.1\,\second$ ({\it left}) and (b) $7.2\,\second$ ({\it right}).
In (a), only the central region of the star is shown.  The radius of
the star is $8 \times 10^{10}\,\cm$. Note the morphological
differences among Models JA (Fig.~\ref{ja}), JB (Fig.~\ref{jb}) and JC
(Fig.~\ref{jc}).
\label{ja}}
\end{figure}

\begin{figure}
\epsscale{1.0}
\plotone{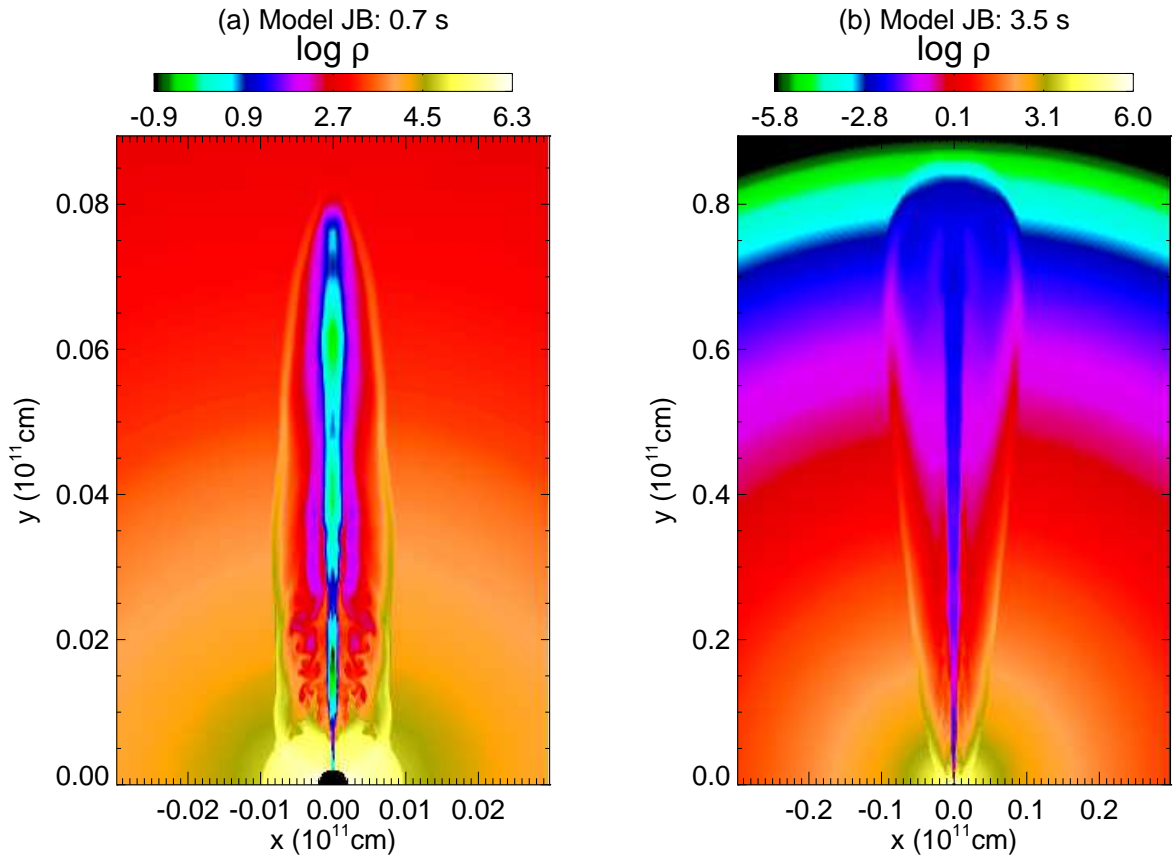}
\caption{Density structure in the local rest frame for Model JB at (a)
$t = 0.7\,\second$ ({\it left}) and (b) $3.5\,\second$ ({\it right}).  In
(a), only the central region of the star is shown.  Note the much
  higher degree of collimation than in Model JA.  
\label{jb}}
\end{figure}

\begin{figure}
\epsscale{1.0}
\plotone{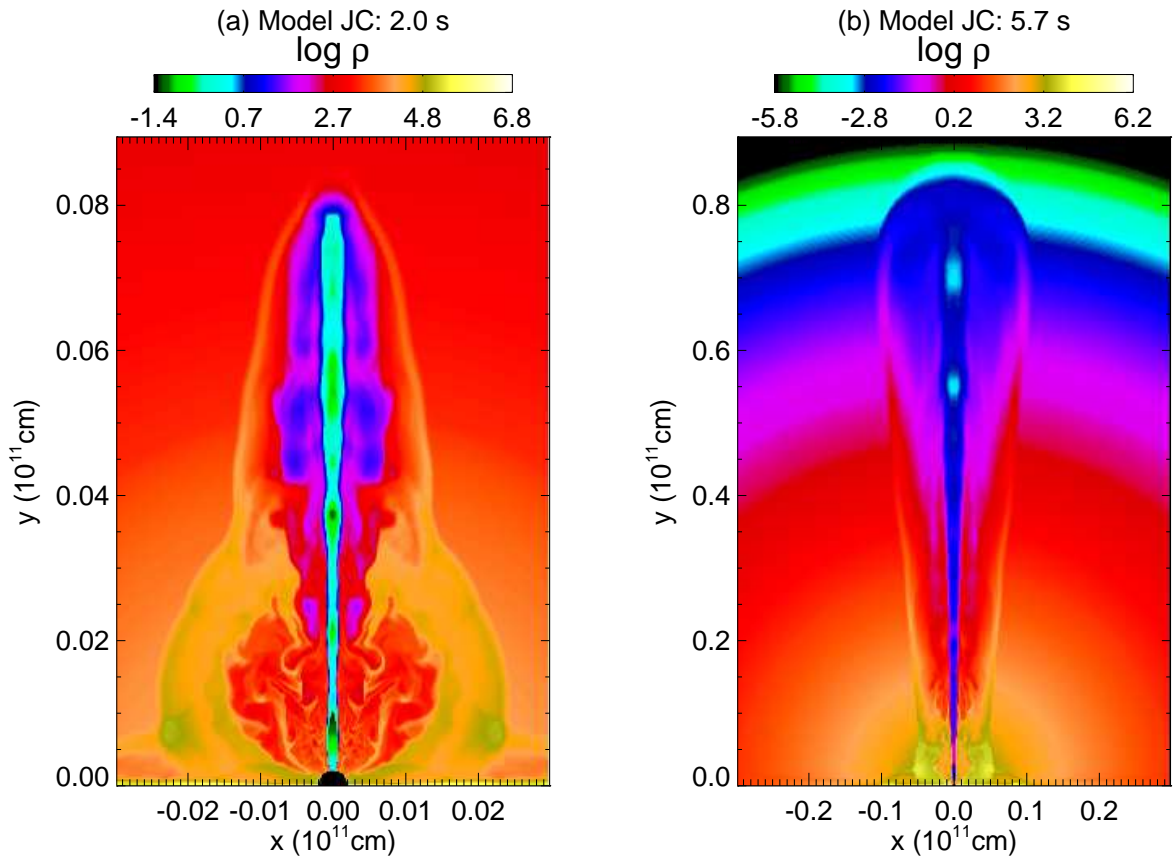}
\caption{Density structure in the local rest frame for Model JC at (a)
$t = 2.0\,\second$ ({\it left}) and (b) $5.7\,\second$ ({\it right}).
In (a), only the central region of the star is shown.  Note the shape
of bow shock in (a) and contrast to Fig.~\ref{ja}.
\label{jc}}
\end{figure}

In all models the jet initially propagates along the polar axis.
After a short time, it consists of a supersonic beam; a cocoon
consisting of shocked jet material and shocked medium gas; a terminal
bow shock; a working surface; and backflows \citep{bla74}.  Some
snapshots of the jet for the three models JA, JB and JC, as defined in
Table~\ref{tab:j}, are shown in Figures~\ref{ja}, \ref{jb} and
\ref{jc}.  In all models, the jet is narrowly collimated and its beam
is very thin, but other aspects of jet morphology vary from model to
model. In Model JA, because of its large initial opening angle, most
of the jet material passes, early on, through a strong shock, which
deflects the flow toward the axis. In Model JB, which has a smaller
initial opening angle and denser beam, the momentum flux is larger
than for Model JA.  The jet head thus advances faster and the beam is
almost naked with a very thin cocoon.  In Model JC, the thickness of the
cocoon and beam and the velocity of the jet head are intermediate
between JA and JB.  At early times, the bow shock has a narrow head
and wide tail (Fig.~\ref{jc}a), reflecting the jet's acceleration.  At
late times however, especially as the jet nears the surface of the
star, the morphologies of all three models are very similar
(Fig.~\ref{ja}b, \ref{jb}b and \ref{jc}b).  This suggests that the
initial conditions tend to be forgotten as the jets propagate and
interact with the star. It also implies that the properties of the
subsequent GRB may not be sensitive to details of the operation of the
central engine.

Because of the ram pressure of infalling stellar material at
$2000\,\km$, a jet cannot start immediately. It must build up some
pressure on the grid.  In Model JA, 0.6 second elapses (in addition to
the $10\,\second$ since core collapse) before the jet starts to
propagate appreciably along the polar axis.  The corresponding delays
are 0.1 second and 0.3 second for Models JB and JC, respectively.
Once they get going, the times for the jets to traverse the star
(outside of $2000\,\km$) and break out also vary: 6.9, 3.3 and 5.5
seconds for Models JA, JB and JC, respectively (Table~\ref{tab:brk}).
This is reasonable since the jet in Model JB has a smaller cross
section area and a larger momentum flux than in Models JA and JC.
 
\begin{figure}
\epsscale{0.45} 
\plotone{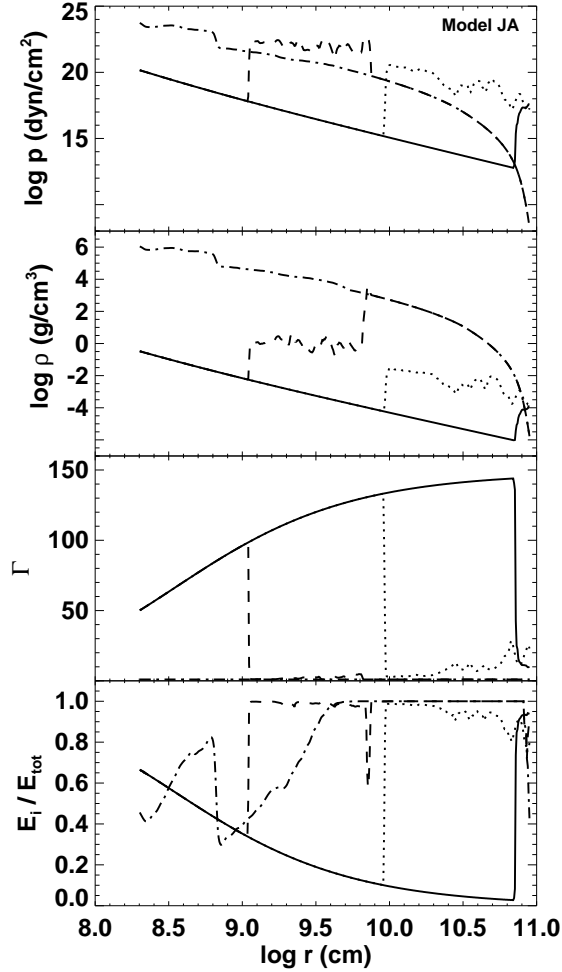}
\caption{Pressure, density, Lorentz factor, and ratio of internal
energy density to total energy density as a function of radius along
the polar axis for Model JA.  Different lines are for different times:
$t=0.0\,\second$ ({\it dash-dotted line}), $t=2.0\,\second$ ({\it
dashed line}), $t=8.0\,\second$ ({\it dotted line}), and
$t=20.0\,\second$ ({\it solid line}).  Pressure and density are
measured in the local rest frame whereas energy density is measured in
the laboratory frame.  Fluctuations of all fluid variables are seen in
the shocked region due to the complex flow patterns (Fig.~\ref{ja}).
In the unshocked region, the flow is steady, reflecting the constancy
of the injected jet.  The total energy density here is the sum of
internal energy and kinetic energy, not including the rest mass
energy.  In the unshocked region, the thermal energy of beam material
is converted into kinetic energy by adiabatic expansion.  At the outer
boundary of the unshocked region, most of the kinetic energy is
converted back into thermal energy by shock dissipation.  At the head
of the jet, the fraction of thermal energy decreases as the unshocked
stellar gas is shocked and its kinetic energy increases.
\label{mulja}}
\end{figure}

\begin{figure}
\epsscale{0.45}
\plotone{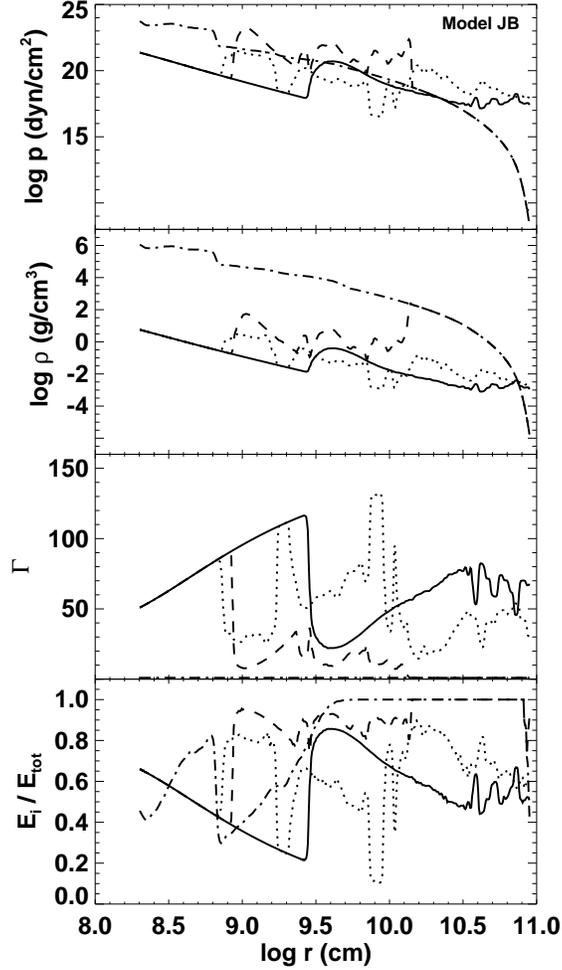}
\caption{Pressure, density, Lorentz factor, and the ratio of internal
energy density to total energy density as a function of radius along
the polar axis for Model JB.  Different lines are for different times:
$t=0.0\,\second$ ({\it dash-dotted line}), $t=1.0\,\second$ ({\it dashed
line}), $t=4.0\,\second$ ({\it dotted line}), and $t=10.0\,\second$
({\it solid line}). The head of unshocked region at $4.0\,\second$
lags behind the head of unshocked region at $1.0\,\second$ because of
mixing among stellar medium, the cocoon, and the jet beam. See
also Figs.~\ref{mulja} and \ref{jb}.
\label{muljb}}
\end{figure}

\begin{figure}
\epsscale{0.45}
\plotone{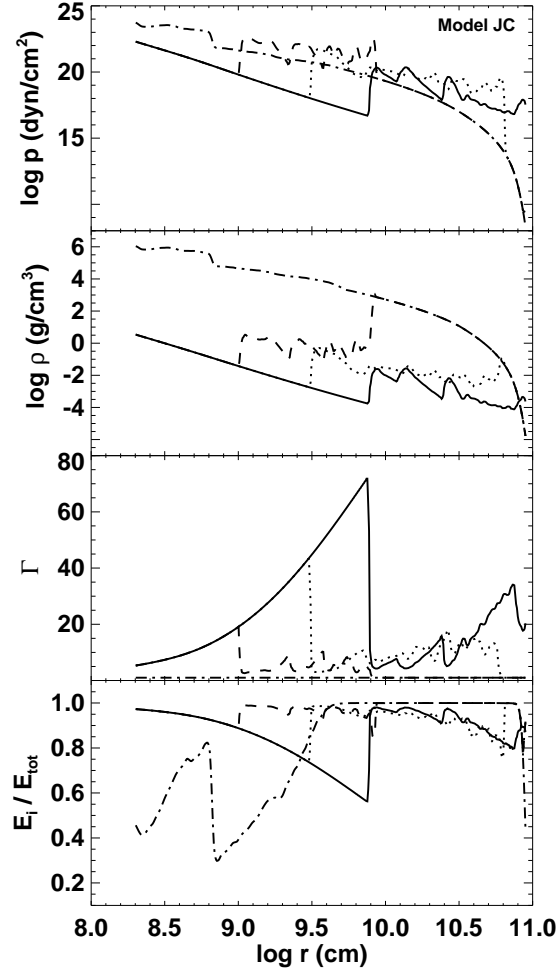}
\caption{Pressure, density, Lorentz factor, and the ratio of internal
energy density to total energy density as a function of radius along
the polar axis for Model JC.  Different lines are for different times:
$t=0.0\,\second$ ({\it dash-dotted line}), $t=2.0\,\second$ ({\it dashed
line}), $t=5.0\,\second$ ({\it dotted line}), and $t=15.0\,\second$
({\it solid line}). See also Figs.~\ref{mulja} and \ref{jc}.
\label{muljc}}
\end{figure}

Along the polar axis, the beam in each model is divided into two
regions: an unshocked region and a shocked region (Fig.~\ref{mulja},
\ref{muljb} and \ref{muljc}).  Because of its low density, low
pressure, and relativistic velocity, the jet initially creates a
low-density, low-pressure funnel along the axis.  Within this funnel,
the jet is steady and expands adiabatically without internal
shocks. The high Lorentz factor helps to suppress the Kelvin-Helmholtz
instability.

Farther out though, the jet head makes its way through the star at a
sub-relativistic speed of roughly $c/2$ (this speed increases as the
jet head moves outward in radius).  Cocoon material flows back from
the working surface. Relativistic material moving along the axis runs
into comparatively stationary stellar matter at the jet head and
pressure grows in this backed up material until it becomes comparable
to that in the cocoon, which is in turn somewhat greater than the
surrounding stellar medium. Then lateral expansion can occur. The need
for deceleration propagates inwards in jet mass - though its location
still moves outward in radius - as a reverse shock. At this inner
shock the adiabatically expanding jet has a ram pressure, $\sim \rho h
\Gamma^2 c^2$, which approximately balances the pressure in the
shocked jet, here $h$, the specific enthalpy, is $(e+p)/(\rho c^2)$,
where, $e$, the total energy density, includes rest mass density.

To summarize, as seen by a piece of jet starting at the origin: 1)
internal energy is converted to expansion until the reverse shock is
encountered; 2) kinetic energy is then converted mostly back into
internal energy at the reverse shock, though the motion remains
moderately relativistic ($\Gamma \sim 5 - 10$); and 3) the hot jet is
further decelerated to subrelativistic speeds at the jet head. In what
follows we shall refer to jet material in stage 1 as the ``unshocked
jet'' and stage 2 as the ``shocked jet''.

The interaction of the shocked jet with the star is especially
interesting since Kelvin-Helmholtz instabilities and oblique shocks
inside the cocoon can imprint time structure on its Lorentz factor
(see also Mart\'{\i} et al. 1997 and references therein). The
consequences differ in the three models.  Because of its larger
initial opening angle, the jet in Model JA experiences a stronger
deceleration at the reverse shock, so the average Lorentz factor of
the shocked jet in Model JA is lower than for Models JB and JC.  In
Model JA, the head of the unshocked jet (i.e., the location of the
reverse shock) moves outward with a gradual acceleration.  In Model JC
the speed is almost constant, but, in Model JB, the location of this
shock even moves backwards sometimes.

\begin{figure}
\epsscale{1.0}
\plotone{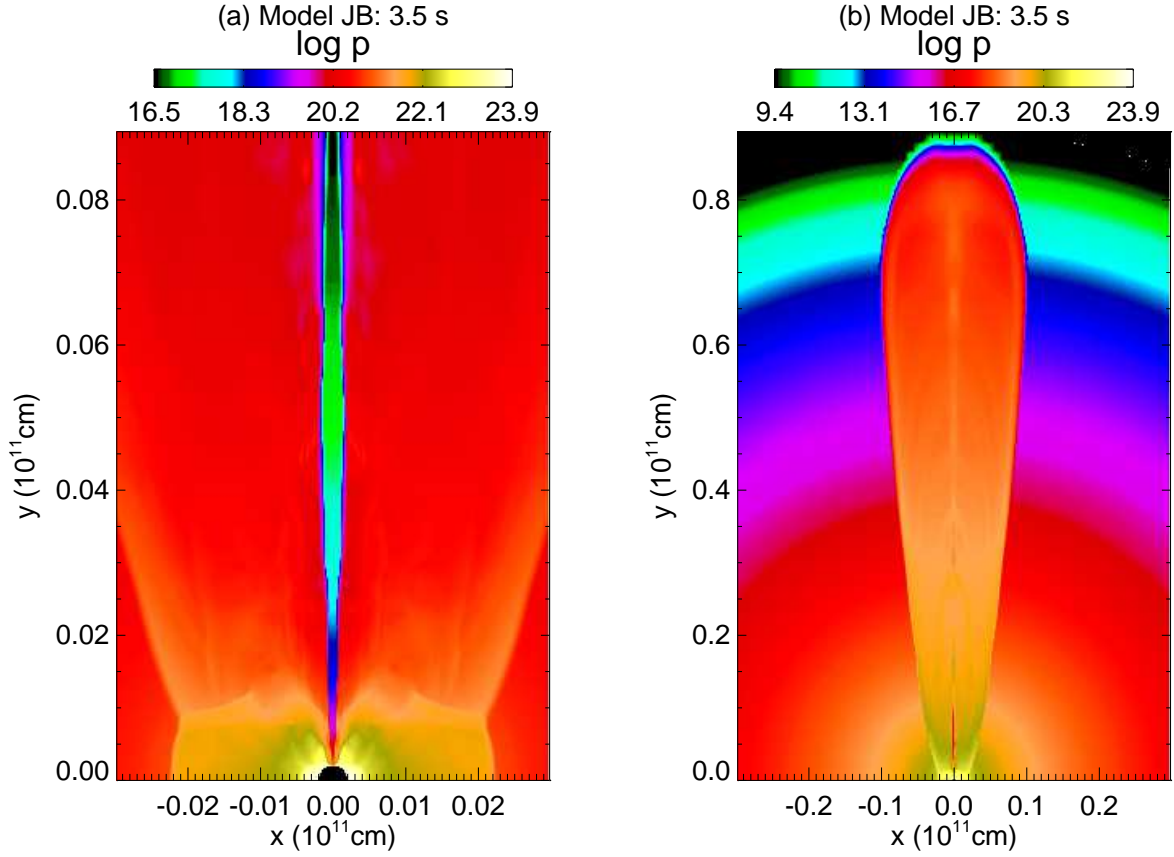}
\caption{Pressure in Model JB at $t=3.5\,\second$.  In (a)
({\it left}) only the central region of the star is shown.  In (b)
({\it right}), the whole star is shown.  The radius of the star is $8
\times 10^{10}\,\cm$.  In the central region, the pressure in the jet
beam is much lower than that in the medium.  The overall structure of
pressure is smooth.
\label{jbpd}}
\end{figure}

A possible explanation is that the unshocked jet in Model JB is
unstable to pinching modes of the Kelvin-Helmholtz instability.  It
has been shown numerically \citep{mar97} and analytically
\citep{fer78} that ultra-relativistic jet beams tend to be unaffected
by the Kelvin-Helmholtz instability.  However, small scale
perturbations might still be able to grow if the jet beam is very thin
\citep{fer78}.  It is also important that these other works assumed
equal pressures in the beam and external medium.  Here the jet is
under-pressurized compared to the medium (Fig.~\ref{jbpd}).  When the
mixing between jet beam and nearly stationary stellar material occurs,
the beam will be decelerated.  This behavior imprints time structure
on the Lorentz factor which consequently turns out to be more variable
Model JB than Models JA and JC.  Some implications of this will be
discussed in \S~\ref{obs}.

\begin{figure}
\epsscale{0.6}
\plotone{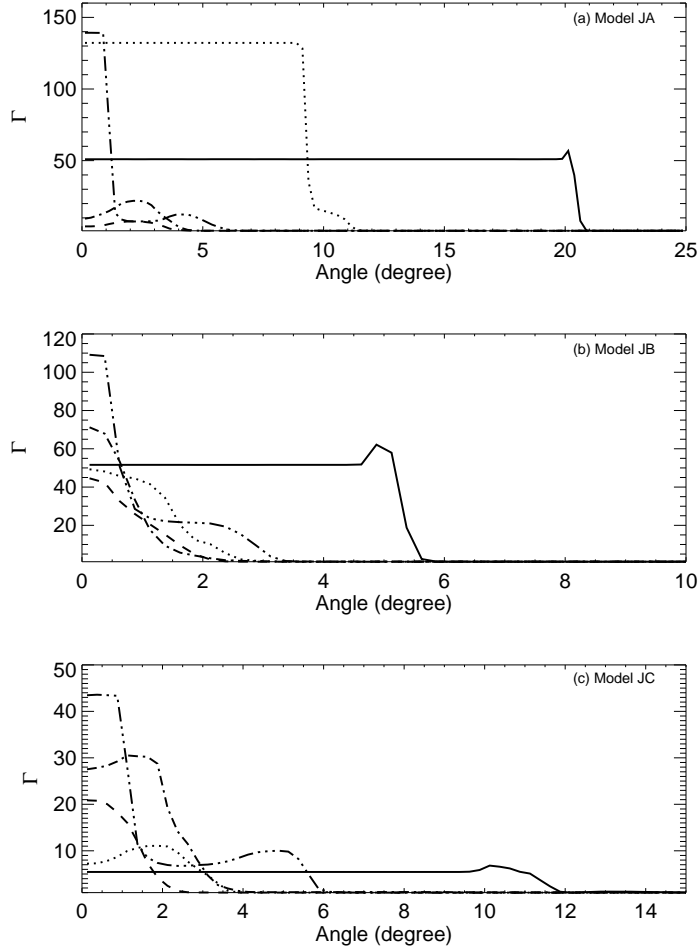}
\caption{Lorentz factor vs. polar angle for: (a) Model JA ({\it top})
at $t=12.0\,\second$, (b) Model JB ({\it middle}) at $t=8.0\,\second$,
and (c) Model JC ({\it bottom}) at $t=10.0\,\second$. These times are
after breakout for the three models. Different lines are for several
radii in the star: $r=2.1\times 10^{8}\,\cm$ ({\it solid lines}),
$r=8.2\times 10^{9}\,\cm$ ({\it dotted lines}), $r=3.1\times
10^{10}\,\cm$ ({\it dashed lines}), and $r=8.0\times 10^{10}\,\cm$
({\it dash dotted lines}).  The dash multi-dotted lines are for: (a)
Model JA ({\it top}) at $r=2.1\times 10^{10}\,\cm$, (b) Model JB ({\it
middle}) at $r=1.8\times 10^{9}\,\cm$, and (c) Model JC ({\it bottom})
at $r=3.0\times 10^{9}\,\cm$. These locations are near the ends of the
unshocked low-pressure and highly relativistic funnels. The jet has a
high Lorentz factor inside the funnel and a mild Lorentz factor
outside.  The radius of the star is $8.0\times 10^{10}\,\cm$.  The
jets are very narrowly beamed as they pass through the star.  The
initial opening angles are $20\degr$, $5\degr$, and $10\degr$ for
Models JA, JB, and JC, respectively.
\label{ang}}
\end{figure}

As the jets pass through the star, they are narrowly collimated by the
external pressure (Fig.~\ref{ang}).  Near the origin, opening angles of
jet beams are close to their initial values, $20\degr$, $5\degr$, and
$10\degr$ for Models JA, JB, and JC, respectively.  In the unshocked
region, the jet beam forms a low-pressure funnel that is quickly
focused by the high pressure of the external stellar matter.  In Model
JA at $t=12.0\,\second$ for example, the half opening angle of the jet
beam has already decreased to $\sim 10\degr$ at $r=8.2 \times
10^{9}\,\cm$, only $1/10$ of radius of the star (Fig.~\ref{ang}).

\begin{figure}
\epsscale{0.7}
\plotone{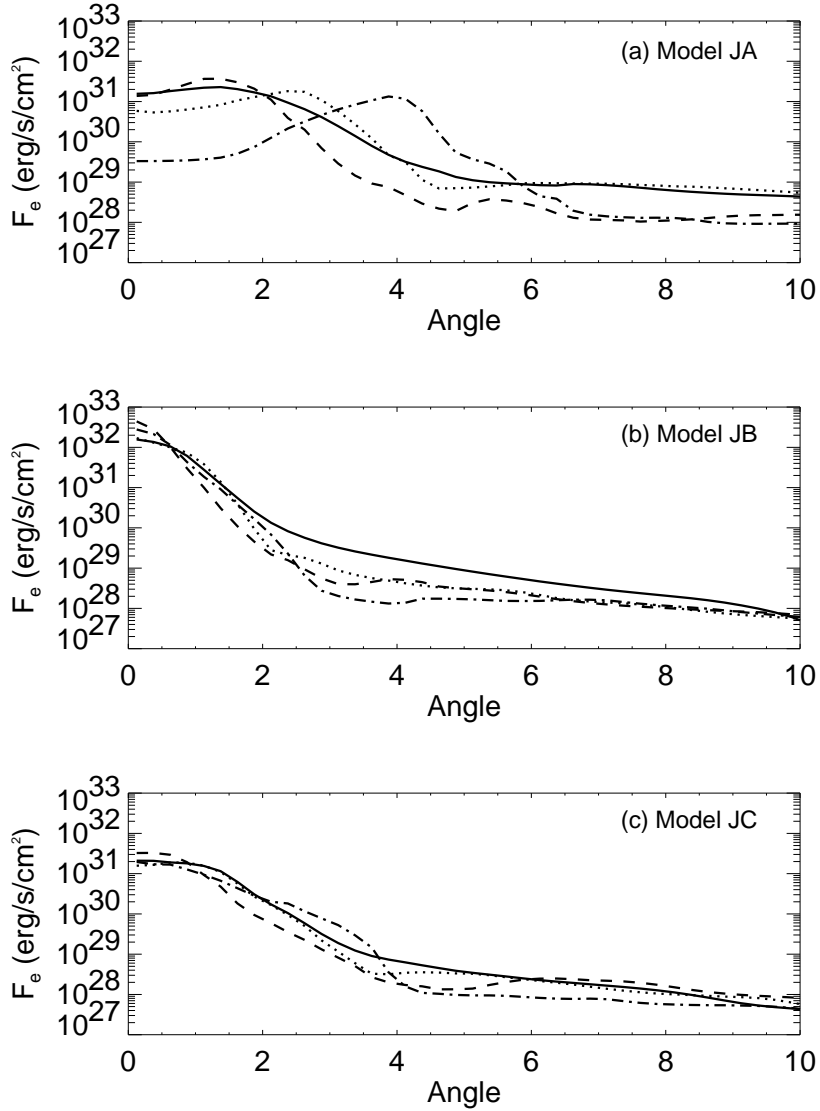}
\caption{Energy flux at $9 \times 10^{10}\,\cm$ for Models JA ({\it
top}), JB ({\it middle}) and JC ({\it bottom}) after breakout.
Here energy means total energy, including both kinetic and
internal.  Different lines are for different times: (a)
$t=8.00\,\second$ ({\it solid line}), $t=12.00\,\second$ ({\it dotted
line}), $t=16.00\,\second$ ({\it dashed line}), and $t=21.85\,\second$
({\it dash dotted line}); (b) $t=4.00\,\second$ ({\it solid line}),
$t=6.00\,\second$ ({\it dotted line}), $t=8.00\,\second$ ({\it
dashed line}), and $t=10.00\,\second$ ({\it dash dotted line}); (c)
$t=6.50\,\second$ ({\it solid line}), $t=9.00\,\second$ ({\it dotted
line}), $t=12.00\,\second$ ({\it dashed line}), and $t=15.00\,\second$
({\it dash dotted line}).  In Model JA, the jet is ``hollow''.  In
Models JB and JC, the energy flux is high in the middle.  The radius
of the star is $8 \times 10^{10}\,\cm$.
\label{fe}}
\end{figure}

In the shocked jet the pressure is larger than in the external medium.
The jet beam is surrounded by an over-pressurized cocoon which is in
turn in balance with the star.  There are no longer dramatic changes
in $\theta_0$ in this shocked region. (Fig.~\ref{ang}).  At the
surface of the star, the half opening angles of the jet beams have
decreased to very similar values, $\sim 4\degr$, $\sim 2\degr$, and
$\sim 3\degr$, for Models JA at $t=12.0\,\second$, JB at
$t=8.0\,\second$, and JC at $t=10.0\,\second$, respectively.  Still
there are interesting differences among the models.  Model JA which
started with the largest initial angle still has the greatest angle at
breakout. On the other hand, the energy flux per solid angle in Model
JB is higher. Thus along the jet it would appear brighter.  The
distribution of energy with angle is also interesting.  In Model JA,
the jet is ``hollow'', but in Models JB and JC, peak flux occurs on
axis.  Fig.~\ref{fe} also shows that the opening angle in Model JA is
increasing as the jet blows off the star. All of these effects,
different intensities per solid angle, more rapidly variable Lorentz
factor in Model JA, changing opening angle with time, etc. will have
important consequences for the GRB light curves and its afterglows.

\subsection{The Emergence of the Jet}\label{breako}

\begin{figure}
\epsscale{0.8}
\plotone{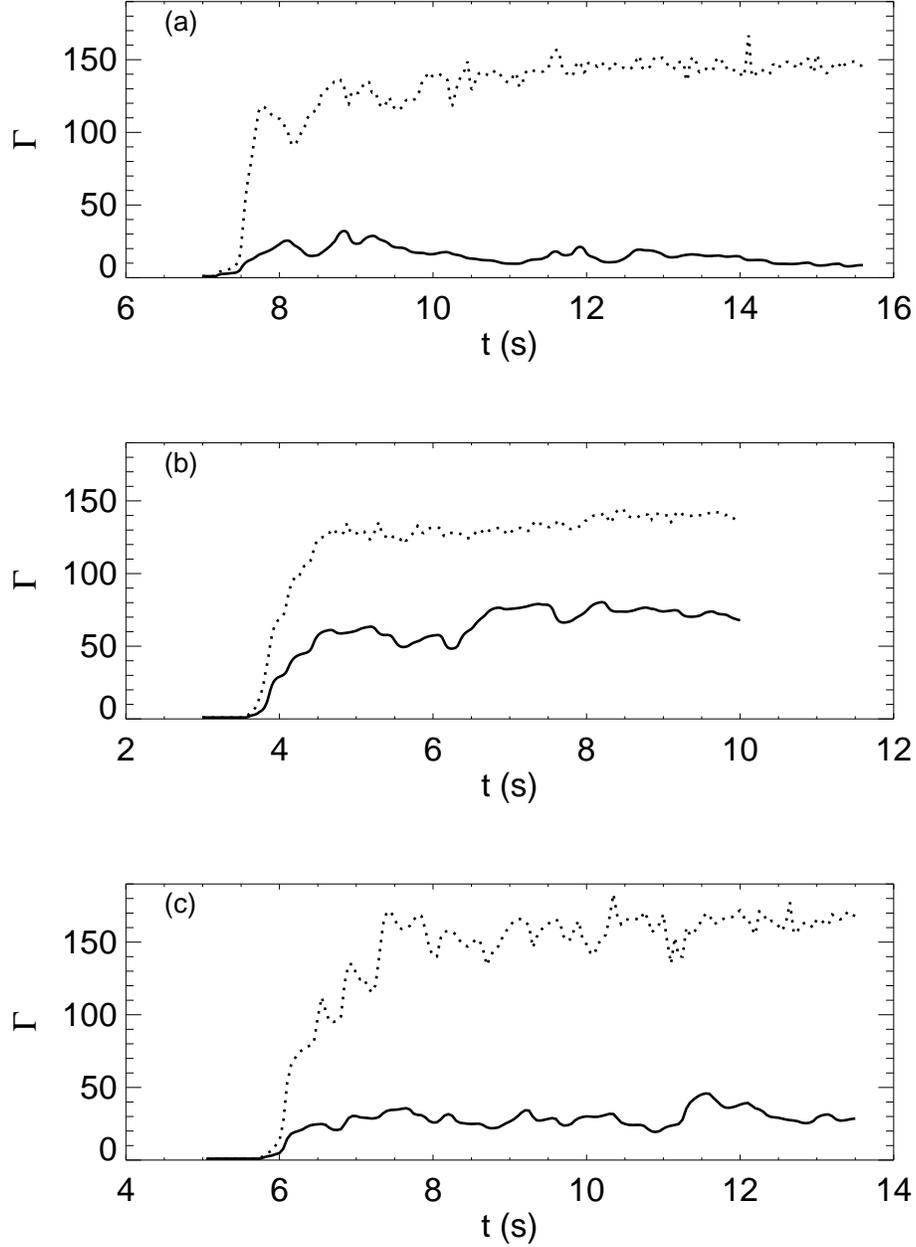}
\caption{Lorentz factor vs. time for: (a) Model JA, (b) Model JB, and
(c) Model JC.  Solid lines show current Lorentz factor of material on
the axis at edge of our computational grid, $r=9\times 10^{11}\,\cm$.
Dotted lines indicate the estimated terminal Lorentz factor if all
internal energy is converted into kinetic energy.
\label{GamT}}
\end{figure}

Eventually, the jet accelerates and breaks free of the star. In Model
JA, the average velocity of the head of the jet just prior to breakout
is $1.2 \times 10^{10}\,\cms$. For Models JB and JC the speeds are
$2.5 \times 10^{10}\,\cms$ and $1.5 \times 10^{10}\,\cms$
respectively. Of course the speeds behind the jet head and in the beam
are higher, $\Gamma \sim$ 10, 50, and 20 for Models JA, JB and JC,
respectively (Fig.~\ref{GamT}). These would be regarded by most as too
low to produce healthy GRBs (e.g., Lithwick \& Sari 2001).  However,
these jets still have a large ratio of internal energy to total
energy: $\sim$ 80-95\%, $\sim$ 50-75\%, and $\sim$ 80-90\% for Models
JA, JB and JC. That is the jets themselves are still fireballs with
$\eta \sim$ 3 to 10. After all internal energy is converted to kinetic
energy, the terminal Lorentz factor will still be $\sim$ 100-200.
This implies that despite being shocked deep inside the star and
experiencing Kelvin-Helmholtz instabilities, the matter that emerges
has almost the same energy per baryon that it started out with at
$2000\,\km$. The baryonic entrainment has been small. Of course this
result needs to be tested in a three dimensional calculation, but it
is encouraging.

\begin{figure}
\epsscale{1.0}
\plotone{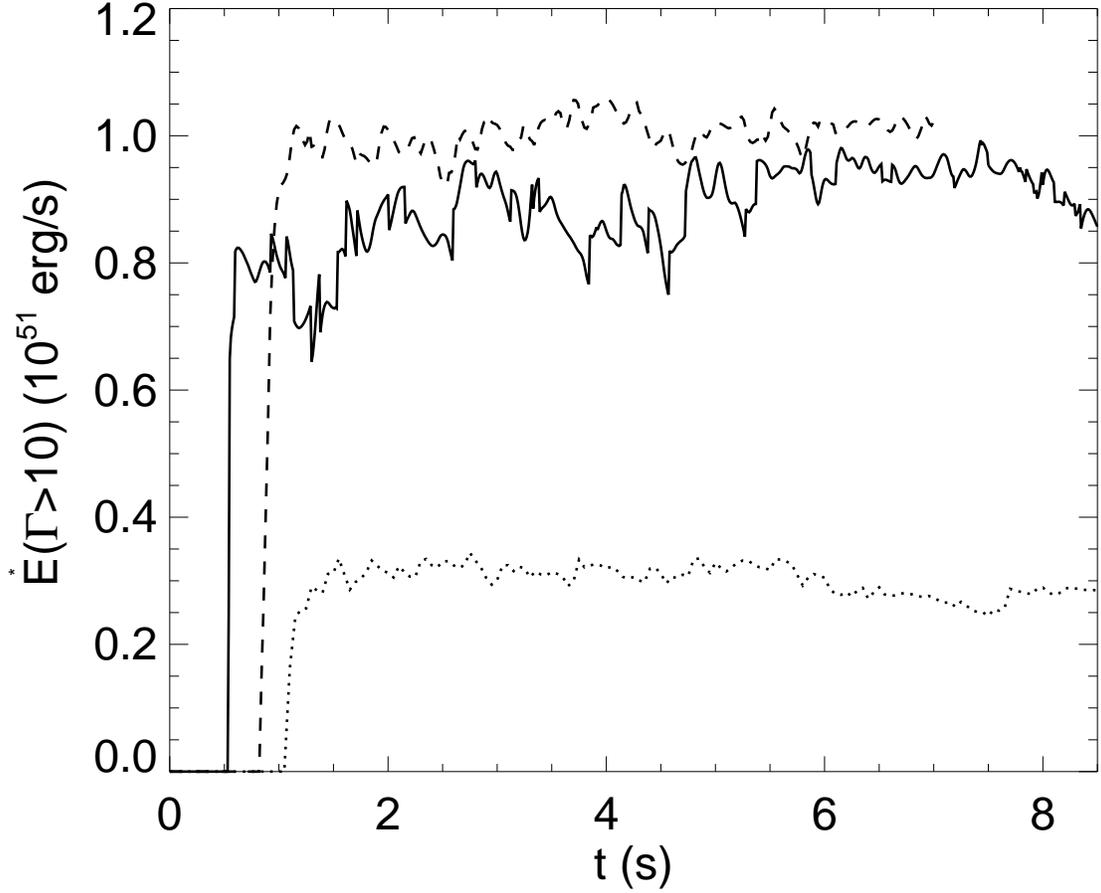}
\caption{Power of the relativistic jets at $9 \times 10^{10}\,\cm$
for: Models JA ({\it solid line}), JB ({\it dashed line}), and JC
({\it dotted line}).  This power includes both kinetic and internal
energy fluxes calculated at the laboratory frame for material which
has a Lorentz factor greater than 10.  Time has been modified to show
the three models together.  The beginning of the solid line for Model
JA is at $7\,\second$; the beginning of the dashed line for Model JB
is at $3\,\second$; the beginning of the dotted line for Model JC is
at $5\,\second$.  Note that the energy flux increases very rapidly
when the jet arrives at the surface. 
\label{edot}}
\end{figure}

After breakout, a channel is open through which the jet can propagate
without losing much energy (though it is still shocked and modulated
while passing through the star). The total power carried by matter
with Lorentz factor greater than 10 is $\sim 0.9 \times
10^{51}\,\ergs$, $\sim 1 \times 10^{51}\,\ergs$ and $\sim 3.0 \times
10^{50}\,\ergs$ just after breakout for Models JA, JB and JC,
respectively (Fig.~\ref{edot}).  Those are very close to the initial
values at the inner boundary (Tab.~\ref{tab:j}).

By the time of breakout, the total energy injected as twin jets was
13.8, 6.6, and $3.3 \times 10^{51}\,\erg$ in Models JA, JB, and JC
respectively. The energy on the grid in various forms is given at that
time in Table~\ref{tab:brk}. In particular we can estimate the
available energy to produce a supernova, $E_\mathrm{sn}$, from the
total energy in matter moving at less than 10\% the speed of light
($\Gamma < 1.005$) by considering corrections due to initial energy
and energy lost at the computational inner boundary.  The jet
propagates very efficiently through the star after breakout
(Fig.~\ref{edot}) so that there is very little further ``wasted''
energy available for powering a supernova in any of the models.
However, one should keep in mind a) that some non-relativistic energy
may still be available to power the supernova as the jet starts to die
off, and b) the wind off the accretion disk may also contribute
significant energy to the explosion (MacFadyen \& Woosley 1999). The
supernova energies in Table~\ref{tab:brk} are thus lower bounds for
the (admittedly high) jet energies assumed.

We followed Models JA, JB, and JC for a total of $21.8\,\second$,
$10.0\,\second$, and $15.0\,\second$ respectively
(Table~\ref{tab:erg}) and their total energies were these times
multiplied by the assumed injection rates (Table~\ref{tab:j}).  Before
breakout some energy is ``wasted'' because the head of the jet moves
much slower than the jet itself.  The amount of ``wasted'' energy is
given by, $E_\mathrm{c} = 2 \dot{E} (t_\mathrm{b} - R_\mathrm{s} /
c)$, where $\dot{E}$ is the power at the base for each jet,
$t_\mathrm{b}$ is the time for the jet to break out of the star, and
$R_\mathrm{s}$ is the stellar radius.  Using this expression, we
estimate the sums of the energies in the supernova and jet cocoon
(Table~\ref{tab:erg}).  Most of this energy is in non-relativistic hot
material near the jet which can power the supernova; the remainder is
in the cocoon which emerges from the star \citep{ram02}.

The history of Lorentz factor at the edge of our computational grid on
the polar axis is shown in Fig.~\ref{GamT}.  This figure also shows
the estimated terminal Lorentz factor if all internal energy is
converted into kinetic energy and the process is adiabatic.  The
fluctuations of the Lorentz factor are not so large as to make
efficient internal shocks, but our grid is relatively coarse and
numerical viscosity may wash out most fluctuations, especially those
due to small scale instabilities.  Because of the large range of radii
covered, our two-dimensional spherical grid consists of
logarithmically spaced radial zones, and both uniform and
logarithmically angular zones.  This gives good resolution at the
center, but unfortunately the resolution near the edge of the star is
too low to study the details of time structure.  The Lorentz factor
along the polar axis at $4.0\,\second$ in Model B contains many
significant fluctuations (Fig.~\ref{muljb}), but these are smoothed
quickly by numerical viscosity.  In the near future
we plan further calculations using cylindrical geometry and finer
resolution to study the mixing better.

When the jet breaks out, it is the hot shocked beam that emerges
first.  If the engine lasts long enough, the unshocked beam may
eventually emerge, e.g. at $\sim 20$ s in Model JA (Fig.~\ref{mulja};
remember that $10\,\second$ must be added to all our times to get the
time when the core actually collapsed).  In Model JB, the mixing
instability caused by the interaction among the jet beam, cocoon and
star slows the jet more. About 6.7 seconds after breakout, the head of
the unshocked jet is still only at at $2.5 \times 10^{9}\,\cm$ while
in Model JC it is at $7.9 \times 10^{9}\,\cm$.  Because more than half
of the total energy in the shocked jet material is in form of internal
energy, it is inevitable that the jet will experience acceleration
and, possibly, sideways expansion.  What is the final opening angle of
the jet?  What is the final Lorentz factor of the jet as the internal
energy is converted into kinetic energy?  In order to power a GRB,
the jet has to have a Lorentz factor of more than 100.  Can the
shocked mildly relativistic jet ($\Gamma \sim 10-50$) acquire high
enough Lorentz factor ($\Gamma > 100$) after breakout?

\subsection{Jet Propagation in the Near Stellar Environment}\label{jout}

After it breaks out of the star, the relativistic jet, which has a
high internal energy loading, accelerates and spreads. It also
encounters the stellar wind surrounding the star. In a second series
of calculations we explored, in a preliminary way, this expansion. The
jet is once more specified by its energy deposition rate, $\dot{E}$,
initial opening angle, $\theta_{0}$, initial Lorentz factor,
$\Gamma_{0}$, and the ratio of kinetic energy to total energy,
$f_{0}$. Two models were considered (Table~\ref{tab:w}): (W1) $\dot{E}
= 8 \times 10^{50}\,\ergs$, $\theta_{0} = 3\degr$, $\Gamma_{0} = 10$,
$f_{0} = 0.06$; (W2) $\dot{E} = 8 \times 10^{50}\,\ergs$, $\theta_{0}
= 3\degr$, $\Gamma_{0} = 50$, $f_{0} = 0.33$.  These conditions are
representative of the actual conditions in the emerging jets for
Models JA, JB, and JC.

Because, for this set of calculations, we are not so interested in
what happens close to the stellar surface, a two-dimensional {\sl
cylindrical} grid ($r, z$) is employed consisting of 2400 zones in
$z$-direction from $10^{11}\,\cm$ to $1.3 \times 10^{12}\,\cm$ and 720
zones in $r$-direction from $0$ to $3.6 \times 10^{11}\,\cm$.  Due to
the relativistic velocities and large radius we can ignore gravity. A
gamma-law equation of state with $\gamma = 4/3$ also suffices. The
background density, which might come from stellar mass loss, has been
assumed to decline as $\rho \sim R^{-2}$, and the density at
$R=10^{11}\,\cm$ is set to $3.7 \times 10^{-11}\,\gcc$, which roughly
corresponds to a mass loss of $\sim 10^{-5}\,\Msunyr$ with a velocity
of $\sim 1000\,\mathrm{km}\,\mathrm{s}^{-1}$.  Actually, the details
of environment are unimportant here because we have not yet followed
the jet to such large radii that the interaction with the wind becomes
important. The jet is injected at $z=10^{11}\,\cm$ with an opening
angle of $3\degr$ with momentum that is purely radial.  Though
potentially quite important, to keep things simple, we do not include
a cocoon in these calculations.  Also for simplicity, the jet is given
uniform initial properties within its opening angle.

\begin{figure}
\epsscale{0.9}
\plotone{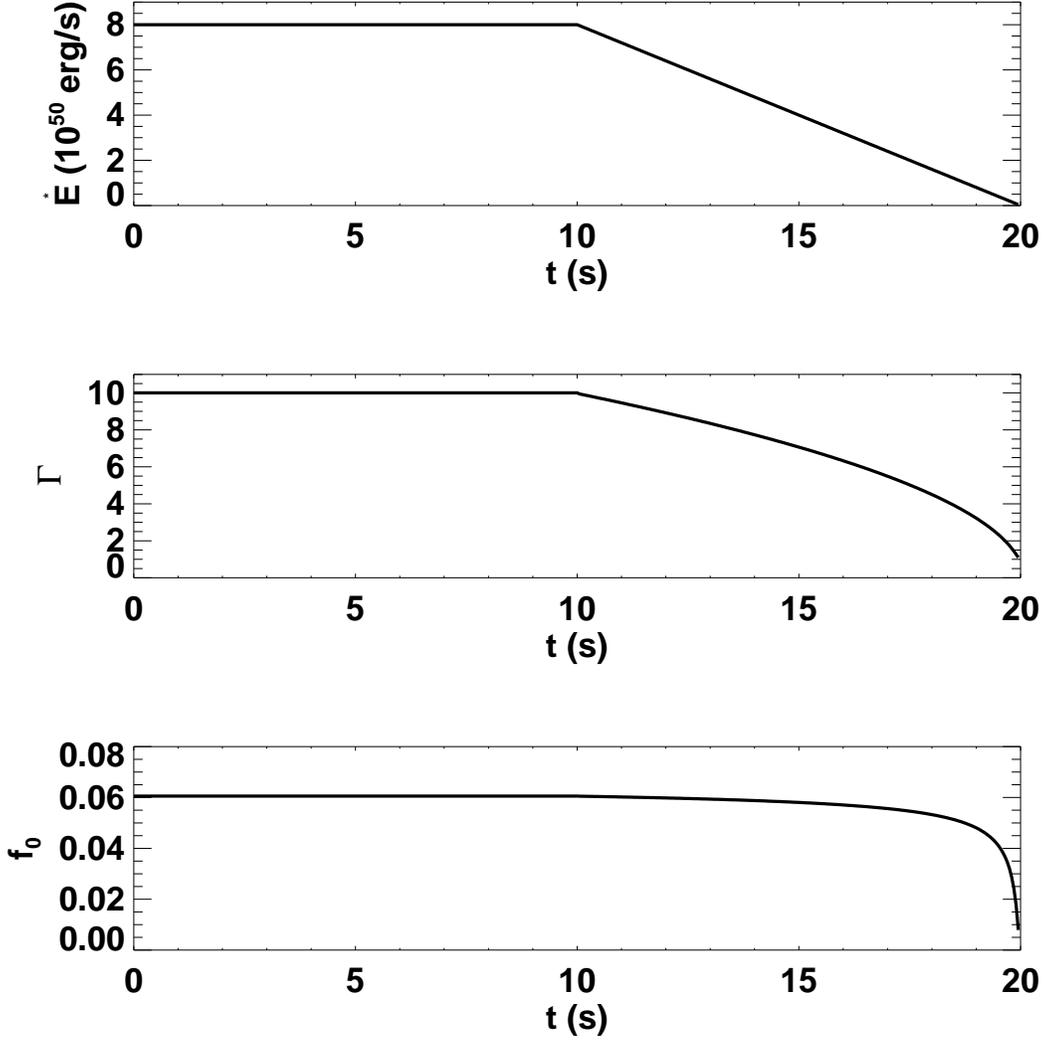}
\caption{Energy deposition rate ({\it top}), Lorentz factor ({\it
middle}), and ratio of kinetic energy density to total energy density
({\it bottom}) of injected jet in Model W1 vs. time. Here, the energy
deposition rate and the total energy density, which are measured in
the laboratory frame, do not include the rest mass energy.  This shows
the history of injected jet in Model W1. During the first 10 seconds,
a constant jet is injected in the stellar wind
(Table~\ref{tab:w}). From 10 to 20 seconds, the jet energy
decays. After 20 seconds, an outflow (zero gradient) boundary
condition is used at the low z boundary.  The recipe for the decaying
jet in Model W2 is similar.
\label{recipe}}
\end{figure}

In our simulations, the jets are injected with the constant properties
specified above for the first 10 seconds.  Eventually though, we
expect the central engine to decline in power and produce a less
energetic jet.  To mimic this decline, the jet energy and Lorentz
factor were both turned down during the interval 10 to $20\,\second$
while the pressure and density of the jet were kept constant.  The
energy deposition rate decreased linearly from $8 \times
10^{50}\,\ergs$ at $ t = 10\,\second$ to 0 at $t=20\,\second$
(Fig.~\ref{recipe}). After 20 seconds, we used an outflow (zero
gradient) boundary condition for the lower $z$ boundary.  The total
energy for each jet is thus $1.2 \times 10^{52}\,\erg$ for both Models
W1 and W2.

\begin{figure}
\epsscale{1.0}
\plotone{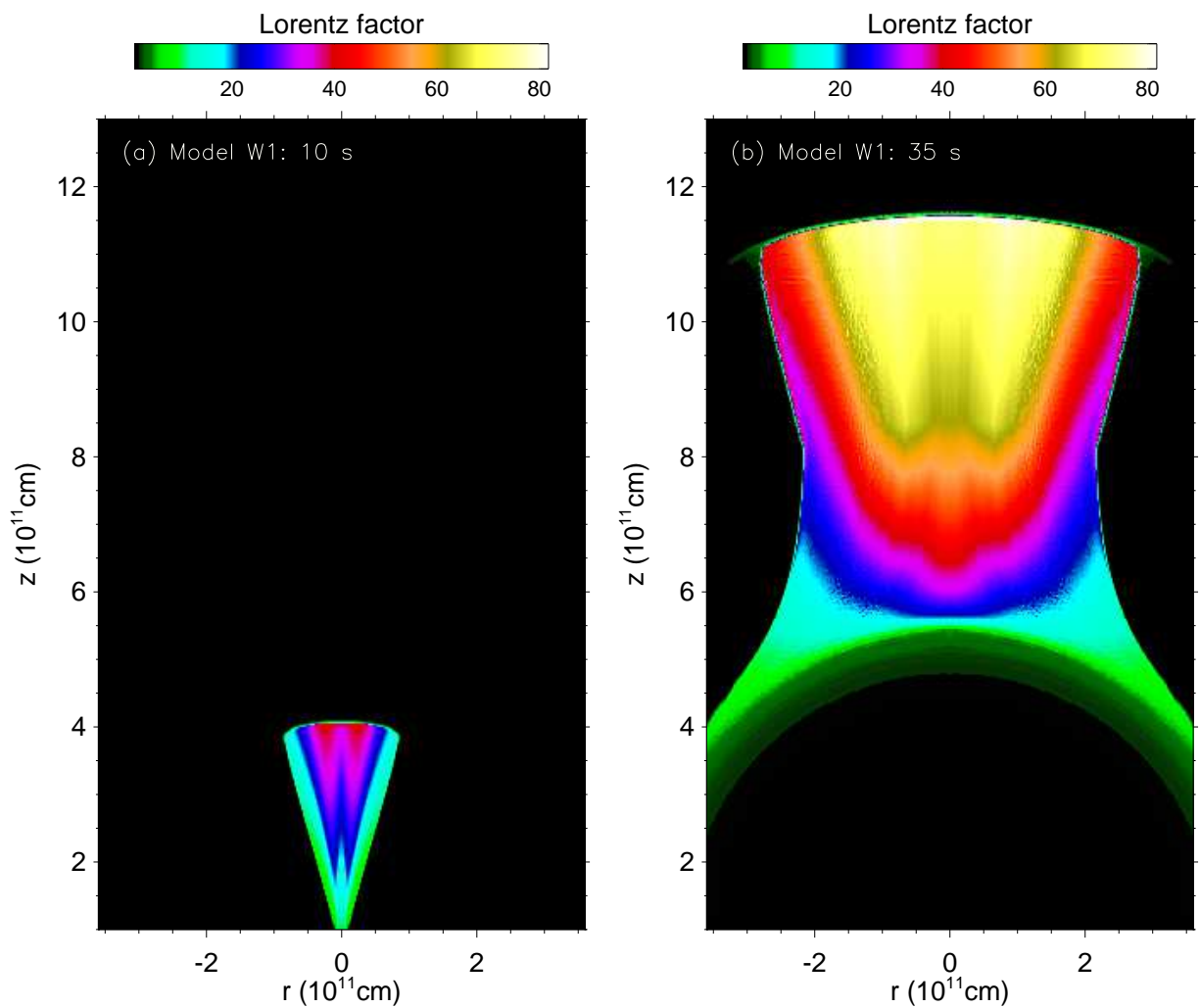}
\caption{Lorentz factor for Model W1 at (a) $t=10\,\second$ ({\it
left}) and (b) $t=35\,\second$ ({\it right}) after the jets start to
propagate in the stellar wind.  The initial opening angle is
$3\degr$. At $t=35\,\second$, the opening angle is $\sim 15\degr$,
which is much bigger than that of Model W2 at the same time.  
Because the power and Lorentz factor decrease gradually after
$10\,\second$, the tail of the jet has much more lateral expansion.
\label{w1}}
\end{figure}

\begin{figure}
\epsscale{1.0}
\plotone{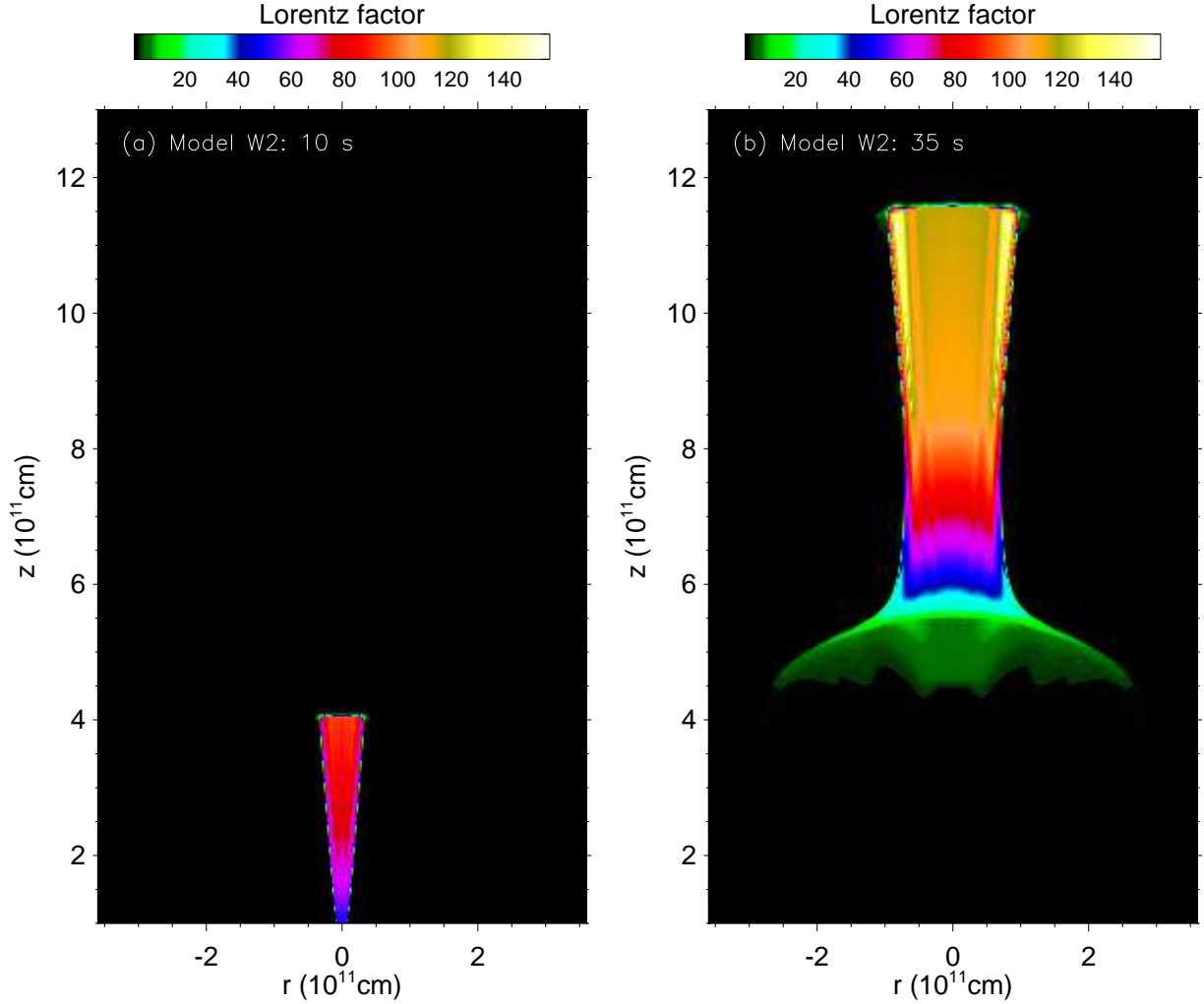}
\caption{Lorentz factor for Model W2 at: (a) $t=10\,\second$ ({\it
left}), and (b) $t=35\,\second$ ({\it right}) after jets start to
propagate in the stellar wind.  The initial opening angle is
$3\degr$. At $t=35\,\second$, the opening angle is $\sim 5\degr$. This
means there is little lateral expansion.  Because the power and
Lorentz factor decrease gradually after $10\,\second$, the tail of
the jet experiences more lateral expansion, though not as much as in
Model W1.
\label{w2}}
\end{figure}

\begin{figure}
\epsscale{1.0} 
\plotone{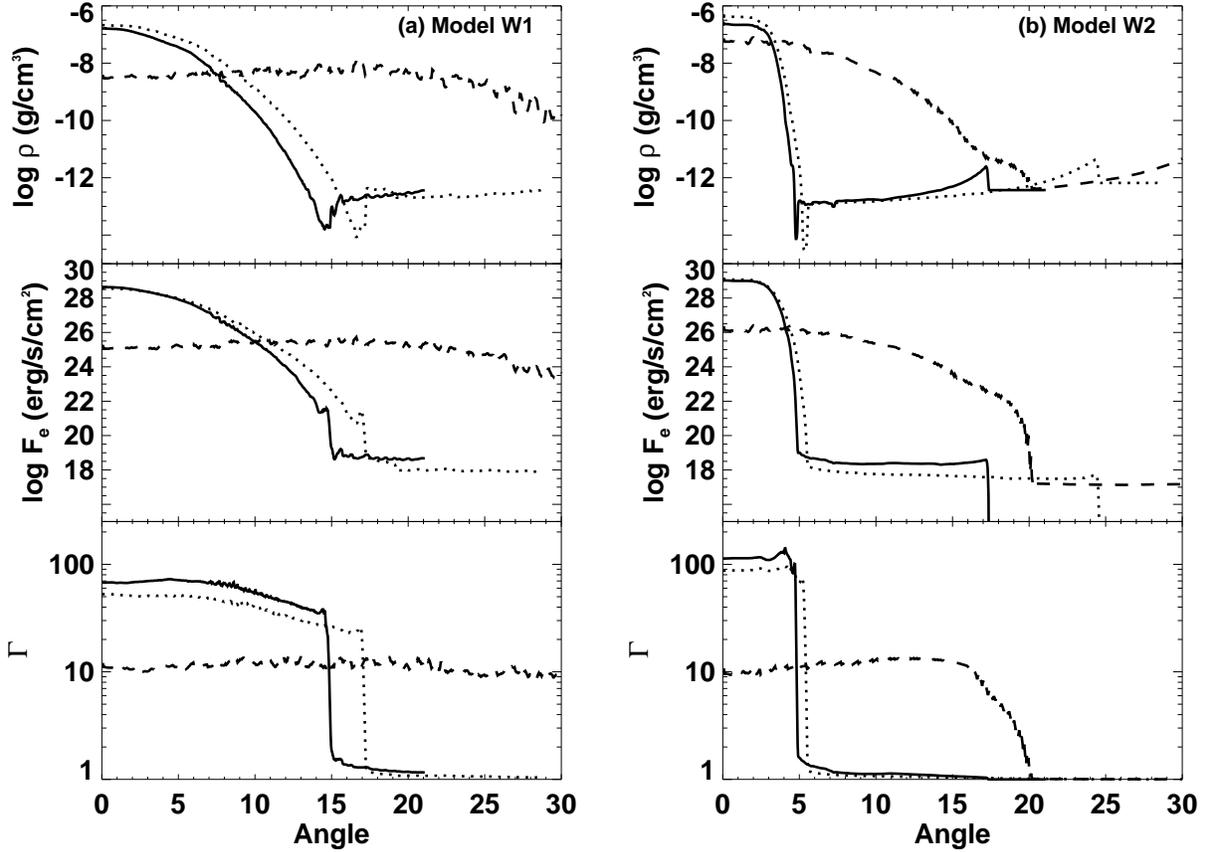}
\caption{Density, energy flux, and Lorentz factor vs. angle for: (a)
Model W1 and (b) W2.  Here ``density'' is the local rest mass density
and ``energy'' is the total energy measured in the lab frame, both
kinetic and internal.  The time is $35\,\second$ after the jets break
out.  Different lines show results at different locations: $r = 5.5
\times 10^{11}\,\cm$ ({\it dashed lines}), $r = 7.5 \times
10^{11}\,\cm$ ({\it dotted lines}), and $r=10^{12}\,\cm$ ({\it solid
lines}).  Here, $r$ is the distance to the center of the star.  The
angle is measured relative to the polar axis starting at the center of
the star.  The opening angles at $t=35\,\second$ are $\sim 15\degr$
and $\sim 5\degr$, for Models W1 and W2, respectively.
\label{fe2}}
\end{figure}

Some results are shown in Figs.~\ref{w1} and \ref{w2}.  In both
models, the Lorentz factor increases to more than 100 as the jet
propagates in the stellar wind.  There are very significant
differences between Models W1 and W2.  Although the initial opening
angle in both is $3\degr$, the opening angle at the head of the jet at
$t=35\,\second$ is $15\degr$ for Model W1 and $5\degr$ for Model W2
(Fig.~\ref{fe2}).  In Model W1 the jet initially had a Lorentz factor
of 10 and a large internal energy (93\% of the total energy) leading
to a large lateral expansion. In Model W2, the internal energy is less
(67\% of the total energy) and the Lorentz factor greater (50).  One
might argue that the final opening angle of the jet should be
$\theta_{0} + 1 / \Gamma_{0}$ (see, e.g., Matzner 2002) in which case
the final opening angles would be $8.7\degr$ and $4.1\degr$ for Models
W1 and W2, respectively.  Our larger angles correctly reflect the
dynamics of a differentially accelerating relativistic jet.

\begin{figure}
\epsscale{1.0} 
\plotone{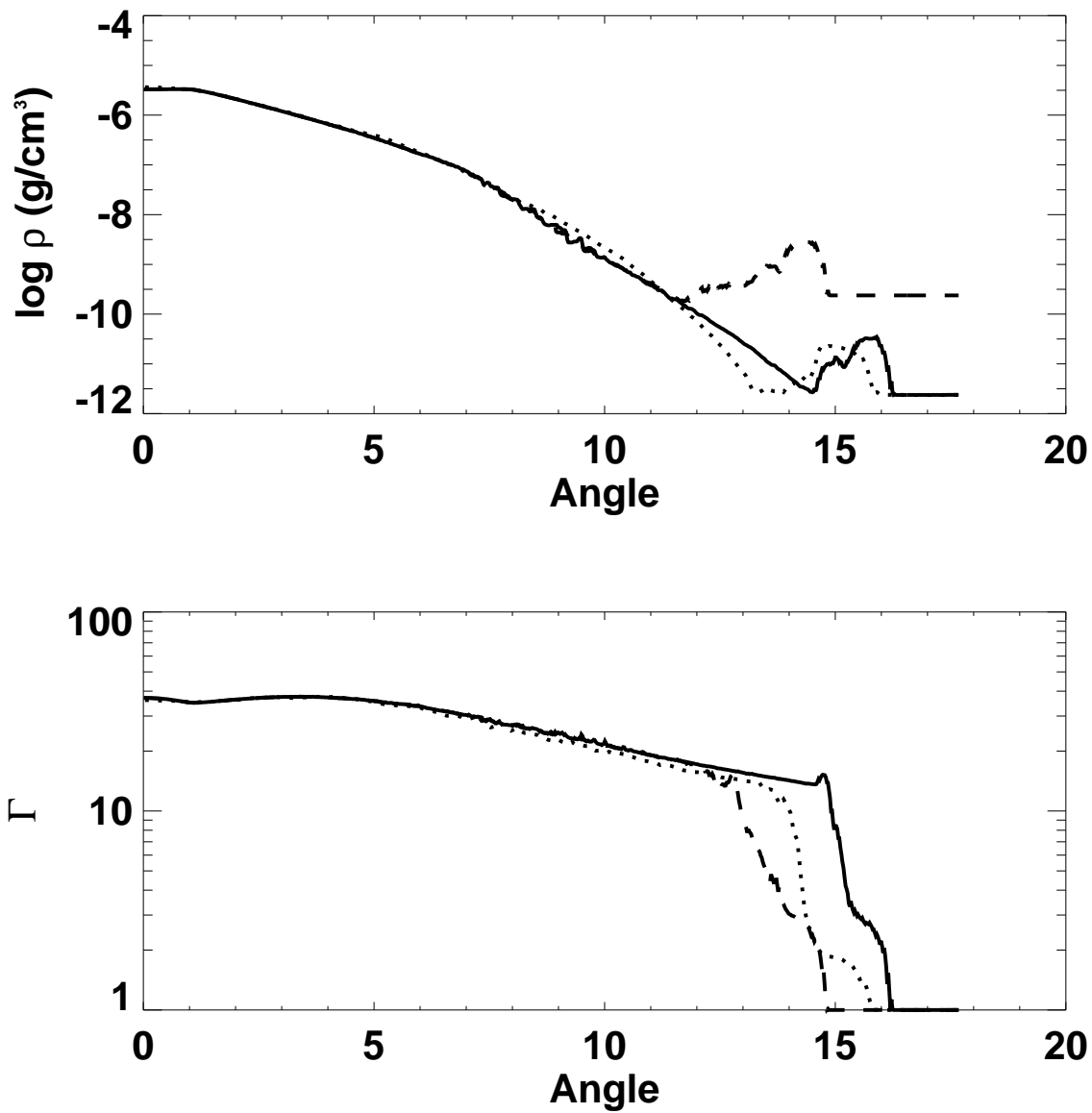}
\caption{ Density and Lorentz factor vs. angle for three calculations
of Model W1 using different resolutions and stellar wind densities.
Here, the density is the local rest mass density. The angle is
measured relative to the polar axis starting at the center of the
star.  The three simulations used: (1) higher resolution and normal
stellar wind density ({\it solid lines}); (2) normal resolution and
normal stellar wind density ({\it dotted lines}); and (3) higher
resolution and higher stellar wind density ({\it dashed lines}). The
time is at $ t=10\,\second$. These lines all show results at $r = 3.95
\times 10^{11}\,\cm$ where $r$ is the distance to the center of the
star.
\label{w1test}}
\end{figure}

As the jet expands sideways, the velocity along its edge does not
increase as rapidly in the radial direction as at the center. A small
amount of material actually ends up, briefly, with a {\sl slower}
radial component.  The conversion of internal energy to kinetic energy
during this phase causes sideways expansion.  It should be noted that
the radial component of momentum still increases when the radial
component of velocity decreases because the total Lorentz factor is
increasing during the sideways expansion.  The combined effect of
these two processes is that a small amount of material ends up at
larger angles than expected from $\theta_{0} + 1 / \Gamma_{0}$.  In
Model W1, because of its lower initial Lorentz factor and higher
internal energy loading, this effect is more pronounced and the
opening angle increases from $3\degr$ to $>6\degr$ during the first
second.  The effect becomes less important as the Lorentz factor
grows.  In contrast, the final opening angle in Model W2 is only
slightly larger than $\theta_{0} + 1 / \Gamma_{0}$ because of its
higher initial Lorentz factor and lower internal energy loading.  In
order to test whether this large sideways expansion is due to low
numerical resolution, we ran two higher resolution tests for Model
W1. Finer grids (up to 3.3 times the original resolution of Model W1)
were run for the same conditions. In one of the tests, the density of
stellar wind was also increased by two orders of magnitude.  These
studies demonstrated that the resolution was already high enough to
ensure the numerical viscosity contributed very little to the sideways
expansion but are somewhat sensitive to the density of stellar wind
(Fig.~\ref{w1test}).  Because of sideways expansion, the energy flux
far off axis in Model W1 is small but not zero. In Model W2, the
profile of energy flux is more uniform and has a sharper edge.  In
both cases the Lorentz factor profile is flat inside the jet in both
models.

At later times, 10 to 20 seconds after the jet was initiated outside
of the star, the energy deposition rate and Lorentz factor were
gradually turned off as described previously.  As a result, the tail
of the jet experiences much more lateral expansion.  Although the
details depend on exactly how the central engine shuts down, it is
natural that the jet should go a period of gradual decline rather than
abruptly shutting off. Fig.~\ref{fe2} shows the energy flux and
Lorentz factor at different locations for the two models.  Mildly
relativistic material is ejected at large angles.  At the end of our
simulations, the jets have two components: a core with high Lorentz
factor and a wide tail with low Lorentz factor.  In Model
W1, the core has a rather flat energy flux profile inside $\sim
3\degr$ and a rapidly decreasing energy flux profile extended to $\sim
15\degr$.  There is also a moderate relativistic ($\Gamma \sim 10$)
tail outside the high Lorentz factor core.  The Lorentz factor of the
core is about $60-100$, but it has not reached its saturated value and
is expected to become $\sim 150$ because of the remaining internal
energy.  In Model W2, the jet has a narrow core of $\sim 5\degr$ with
a very flat energy flux profile. It also contains moderately
relativistic material outside the high Lorentz factor core. The
implications for GRBs will now be discussed.

\section{OBSERVATIONAL IMPLICATIONS}
\label{obs}

\subsection{Beaming}

Our special relativistic calculations show, as did \citet{alo00}, that
a jet originating near the center of a collapsing massive star will
emerge with an opening angle $\lesssim 5$ degrees. This result is
relatively insensitive to the initial opening angle. Further expansion
occurs after the jet has already emerged from the star. Depending on
its Lorentz factor and internal energy loading at breakout, the final
opening can be smaller than 5 degrees, or more than 15 degrees. This
suggests that gamma-ray bursts can have a wide range of opening angle,
and can be beamed to less than 1\% of the sky with obvious
implications for the requisite GRB energy and event rate.

\subsection{GRB Light Curves}

For a typical total energy in relativistic matter ($\Gamma >$ a few)
of $\sim 3 \times 10^{51}\,\erg$ \citep{fra01, fre01}, and an assumed
Lorentz factor, $\Gamma \gtrsim 100$ \citep{lit01}, the mass within
the GRB-producing beam is $\sim 10^{-5}\,\Msun$, or about 1000 times
less than that within the same solid angle in the precollapse star. In
order to make a GRB using a jet launched from the stellar center, this
stellar matter must be pushed aside. The matter shoved aside has two
consequences. First, its lateral motion initiates a shock wave that
moves round the star converging on the equator and exploding at least
the outer part of the star as a supernova. Second, some of the
material is swept back creating the cocoon of the jet. Instabilities
between the cocoon and jet beam modulate the flow leading to a
variable Lorentz factor. It is even possible that the jet might be
temporarily clogged and reopened by the flow.

In this fashion, the variable Lorentz factor needed for the internal
shock model for GRBs is generated. Light curves having very complex
temporal behavior can result in situations where the central engine
produced only a jet with constant power. The converse is also true.
Short term time structure (less than the jet transit time through the
star or about $5\,\second$) of the central engine may be erased as the
jet passes through the reverse shock. If the variability of the engine
itself is visible at all, it may only be at very late times when the
star has exploded.

Our calculations show significant variation in the Lorentz factor of
the emergent jet (Figs.~\ref{mulja}, \ref{muljb}, \ref{muljc} and
\ref{GamT}). We suspect these variations may be larger in calculations
with higher resolution, and in three dimensions.

Our model also suggests that the variability of GRB light curves may
evolve with time. Though it takes awhile, of order a minute, for the
star to explode away from the jet opening its channel appreciably, the
Kelvin-Helmholtz instabilities should decrease with time. This implies
that GRB variability may also decrease with time.  A narrower jet
experiences more instabilities and therefore the GRB is more
variable. For a given total energy input at the bottom, a narrower jet
also carries more energy per solid angle. Perhaps this might help to
explain the observed correlation between luminosity and variability
\citep{fen00,rei01}.

\subsection{A Unified Model For High Energy Transients}
\label{unified}

According to the ``Unified Model'' for active galactic nuclei (e.g.,
Antonucci 1993), one sees a variety of phenomena depending upon the
angle at which a standard source is viewed. These range from
tremendously luminous blazars, thought to be jets seen on axis, to
narrow line radio galaxies and Type 2 Seyferts thought to be similar
sources seen edge on. Given that an accreting black hole and
relativistic jet may be involved in both, it is natural to seek
analogies with GRBs.

In the equatorial plane of a collapsar - the common case - probably
little more is seen than an extraordinary supernova. Along the axis,
one sees an ordinary GRB, but there may also be interesting phenomena
at intermediate angles. The calculations presented here clearly show
that the edges of jets are not discontinuous surfaces. Moving off
axis, one expects and calculates a smooth decline in the Lorentz
factor and energy of relativistic ejecta. These low energy wings with
moderate Lorentz factor come about in four ways. First, the jet that
breaks out still has a lot of internal energy. Expansion of this
material in the co-moving frame leads to a broadening of the jet. As a
result a small amount of material with low energy ends up moving with
intermediate Lorentz factors - say 10 - 30 and at angles up to several
times that of the main GRB-producing jet. Second, as the star explodes
from around the jet, the emerging beam opens up. Even at constant
total power, the power per solid angle will decline as this channel
broadens. Third, the jet is surrounded by a hot mildly relativistic
cocoon. This material has low energy and low Lorentz factor. It can
expand to large angles. Fourth, and potentially most important, the
jet can continue with a declining total power for a long time.  It is
natural that the Lorentz factor of the emerging jet decline as well. A
hot jet emerging with lower Lorentz factor will expand to larger
angles (\S~\ref{jout}).  As a consequence of the spreading of the jet,
a large region of the sky, much larger than that which sees the main
GRB, will see a hard transient with less power, lower Lorentz factor,
and perhaps coming from an external shock instead of internal ones.

We have speculated for some time now \citep{woo99a,woo99b,woo00,woo01}
that these ordinary bursts seen off axis might appear as hard x-ray
transients of one sort or another. We have identified them with GRB
980425 and with the class of hard x-ray flashes reported by
\citet{hei01}. These are not ordinary high $\Gamma$ jets seen just
beyond a sharp edge \citep{iok01}. The events are made by matter
moving toward us.

\subsection{Breakout Transients and Short Hard Bursts}
\label{shb}

The breakout of the jet and its interaction with the stellar wind will
surely lead to some sort of ``precursor'' event. In fact, it is
possible that this event could, in some cases, dominate the display.

The total energy in a short hard GRB (SHB) is, on the average, about
1/50 that of a long-soft burst (LSB). If the total gamma-ray energy of
the latter averages $3 \times 10^{50}\,\erg$, then SHBs at the same
distance have gamma-ray energy, including the effects of beaming, of
$\sim 10^{49}\,\erg$, or $\sim 3 \times 10^{51}\,\erg$ in equivalent
isotropic energy.

Our present calculations lack adequate resolution at the surface of
the star to follow shock breakout with any accuracy. They also have
not been followed sufficiently long to see the interaction of the
leading head of the jet with the circumstellar medium.  However, we do
focus approximately $5 \times 10^{51}\,\erg$ on only 1/800 of the
stellar surface (for a fiducial opening half-angle of 4 degrees and a
jet that takes $5\,\second$ to exit the star). This corresponds to an
equivalent isotropic energy of $\sim 4 \times 10^{54}\,\erg$. For
their fiducial model with total isotropic energy $5 \times
10^{54}\,\erg$, \citet{tan01} calculate about $3 \times 10^{51}\,\erg$
is accelerated to Lorentz factor $\Gamma > 40$.  It may be that a
larger amount of material is carried along by the leading edge of the
jet as it expands and accelerates outside the star (see also Waxman \&
M\'{e}sz\'{a}ros 2002) if the lateral expansion of the ``plug'' cannot
keep pace with the radial expansion of the jet.

Material having $3 \times 10^{51}\,\erg$ in kinetic energy and $\Gamma
\sim 40$, has a mass $\sim 4 \times 10^{-5}$ $\Msun$ which will
decelerate after running into $10^{-6}\,\Msun$. If the mass loss rate
of the star during its last days was $10^{-4}\,\Msunyr$ and the
velocity $10^8\,\cms$, this deceleration will occur within $3 \times
10^{13}\,\cm$ or $1000\,\second$ in the lab frame. The duration of the
burst is shortened by $\Gamma^{-2}$ to be less than a second, that is
it may be a short one.  But why then don't we see such short hard
precursors in all GRBs and how can one get short hard bursts in
isolation?

The absence of SHBs in LSBs could, to some extent, be semantic, If the
burst continues, a spike at the beginning may be counted as part of a
long burst. But what about the converse problem - SHBs seen in
isolation? The relative brightness of the breakout burst to the
subsequent burst depends on many factors - the circumstellar density,
the total energy of the jet, the variation of its Lorentz factor
(especially critical in the internal shock model), and the mass of the
star. This is especially so since SHBs could be made by external
shocks and LSBs by internal shocks. So it may be, occasionally, that
the breakout burst is an order of magnitude brighter (in flux, if not
fluence) than the ensuing long burst. Of course, not all SHBs need to
be produced this way and other models like merging neutron stars
remain viable.

This possibility suggests that a careful search for postburst activity
might be productive in SHBs. Preliminary results by \citet{con02} are
suggestive of continuing postburst emission on a time scale not unlike
that of LSBs, but are not sufficiently sensitive for a detailed
comparison.

\subsection{Post-burst Phenomenology}

After the main burst is over, accretion will continue at a decaying
rate. The lateral shock launched by the jet starts at the pole and
wraps around the star, but does not reach into the origin at the
equator. One may envision an angle-dependent ``mass
cut''. Consequently, some reservoir of matter remains to be accreted
at late time. This accretion occurs at a rate governed by the
viscosity of the residual disk and the free fall time of material
farther out not ejected in the supernova. \citet{mac01} and
\citet{che89} estimate the accretion rate from fall back to be $\sim3
\times 10^{-6} t_4^{-5/3}\,\Msun\,\second^{-1}$. Here $t_4$ is the
elapsed time since core collapse in units of $10^4\,\second$. Given
the slow rate, the disk that forms is not neutrino dominated and there
may be considerable high velocity flow from its surface
\citep{mac99,nar01}.  That outflow will still be jet-like in nature
since the equatorial plane is blocked by the disk and its energy will
be $\sim 5 \times 10^{46} t_4^{-5/3} \epsilon_{-2}\,\ergs$ where
$\epsilon_{-2} = \epsilon/10^{-2}$ is the efficiency for converting
rest mass into outflow kinetic energy measured in percent. This is
comparable to the energy in x-ray afterglows and might be important
for producing the emission lines reported in some bursts \citep{ree00,
mes01, mcl02} and for providing an extended tail of hard emission in
the GRB itself.

As a consequence of this continuing outflow, the polar regions of the
supernova made by the GRB remain evacuated and the photosphere of the
object resembles an ellipse seen along its major axis but with conical
sections removed along the axis. An observer can see deeper into the
explosion than they could have without the operation of the jet's
``afterburner''.

\section{FUTURE WORK}

The present study, and its precursors by \citet{mac99} and
\citet{alo00}, have shown the critical role of the jet-star
interaction in producing the observable properties of GRBs.  Further
work will be needed though to confirm the validity of the present
results and add strength to some of the model's more speculative
aspects (\S~\ref{obs}).

First, it is important to repeat our calculations in three
dimensions. Use of two dimensions imposes a rotational symmetry that
makes it difficult for instabilities to deflect the beam. It may be
that jets which maintain a stable focus here in 2D will waiver and
even disperse in 3D. In any case, the Kelvin-Helmholtz instabilities
may behave differently. Fortunately the computational requirements are
not too extreme and such a calculation can be carried out in the near
future.

Second, our jet needs to be studied with higher resolution. One can do
that using cylindrical coordinates and a smaller range of
radii. Again, we need to know the nature of the relativistic
Kelvin-Helmholtz instability better.

The emergence of the jet and its immediate interaction with the
circumstellar medium needs further examination, especially in a
calculation that finally resolves the stellar surface at breakout and
includes the jet cocoon as well as its core (see also Ramirez-Ruiz et
al. 2002).

Finally, we are interested in the long term evolution of the jet and
the star it is exploding. How much of the star remains to accrete even
after the initial shock from the jet has wrapped around the star? How
will the object appear, not seconds, but hours and days after the
burst?  How does this relate to the possibility of x-ray lines being
detected in GRB afterglows.

Calculations to address all of these issues are underway.

\acknowledgments

We appreciate many helpful conversations on the subject of GRBs with
Chris Fryer, Alex Heger, Chryssa Kouveliotou, Chris Matzner, and
Martin Rees and are particularly indebted to Dr. Heger for providing
the unpublished precollapse model used in this calculation.  We thank
Enrico Ramirez-Ruiz and M. A. Aloy for helpful comments on the
manuscript.  We also thank the anonymous referee for useful
comments. This research has been supported by NASA (NAG5-8128,
NAG5-12036, and MIT-292701) and the DOE Program for Scientific
Discovery through Advanced Computing (SciDAC; DE-FC02-01ER41176).

\clearpage

\begin{deluxetable}{ccccc}
\tablecaption{Parameters of Models for Jet Propagation inside Star 
              \label{tab:j}}
\tablewidth{0pt}
\tablehead{
\colhead{} &
\colhead{$\dot{E}$\tablenotemark{a}} &
\colhead{$\theta_{0}$\tablenotemark{b}} &
\colhead{} &
\colhead{} \\
\colhead{Model} &
\colhead{($10^{51}\,\ergs$)} &
\colhead{(degree)} &
\colhead{$\Gamma_{0}$\tablenotemark{c}} &
\colhead{$f_{0}$\tablenotemark{d}} 
}
\startdata
 JA & 1.0 & 20 & 50 &  0.33 \\
 JB & 1.0 &  5 & 50 &  0.33 \\
 JC & 0.3 & 10 &  5 & 0.025 
\enddata
\tablenotetext{a}{Energy deposition rate for each jet}
\tablenotetext{b}{Initial half angle}
\tablenotetext{c}{Initial Lorentz factor}
\tablenotetext{d}{Initial ratio of kinetic energy to total
  energy (excluding the rest mass)}
\end{deluxetable}

\clearpage

\begin{deluxetable}{cccccccc}
\tablecaption{Energies at Breakout \label{tab:brk}} 
\tablewidth{0pt} 
\tablehead{ 
\colhead{} &
\colhead{$t_\mathrm{b}$\tablenotemark{a}} &
\colhead{$E_\mathrm{inj}$\tablenotemark{b}} & 
\colhead{$E_{1}$\tablenotemark{c}} & 
\colhead{$E_{2}$\tablenotemark{d}} &
\colhead{$E_{3}$\tablenotemark{e}} &
\colhead{$E_{4}$\tablenotemark{f}} &
\colhead{$E_\mathrm{sn}$\tablenotemark{g}} \\
\colhead{Model} & 
\colhead{($\second$)} & 
\colhead{($10^{51}\,\erg$)} &
\colhead{($10^{51}\,\erg$)} &
\colhead{($10^{51}\,\erg$)} &
\colhead{($10^{51}\,\erg$)} &
\colhead{($10^{51}\,\erg$)} &
\colhead{($10^{51}\,\erg$)} 
}
\startdata 
JA & 6.9 & 13.8 & 9.8 & 2.6 & 0.7 & 4.3 & 6.2 \\ 
JB & 3.3 &  6.6 & 3.6 & 0.4 & 0.1 & 4.9 & 1.2 \\ 
JC & 5.5 &  3.3 & 7.4 & 0.3 & 0.2 & 1.6 & 1.2
\enddata
\tablenotetext{a}{Time for the jet to break out of the star} 
\tablenotetext{b}{Total energy injected as twin jets before the jet
  breaks out of the star}
\tablenotetext{c}{Total energy in material with a Lorentz of 
  $\Gamma <  1.005$}
\tablenotetext{d}{Total energy in material with a Lorentz of 
  $1.005 < \Gamma < 2$}
\tablenotetext{e}{Total energy in material with a Lorentz of 
  $2 < \Gamma < 5$}
\tablenotetext{f}{Total energy in material with a Lorentz of 
  $\Gamma > 5$}
\tablenotetext{g}{Estimated energy in supernova (see \S~\ref{breako})}
\end{deluxetable}

\clearpage

\begin{deluxetable}{ccccc}
\tablecaption{Total Energies at End of Computation \label{tab:erg}} 
\tablewidth{0pt} 
\tablehead{ 
\colhead{} &
\colhead{$T$\tablenotemark{a}} &
\colhead{$E_\mathrm{t}$\tablenotemark{b}} & 
\colhead{$E_\mathrm{w}$\tablenotemark{c}} &
\colhead{$E_\mathrm{j}$\tablenotemark{d}} \\
\colhead{Model} & 
\colhead{($\second$)} & 
\colhead{($10^{51}\,\erg$)} &
\colhead{($10^{51}\,\erg$)} &
\colhead{($10^{51}\,\erg$)} 
}
\startdata 
JA & 21.8 & 43.6 & 8.5 & 35.1 \\
JB & 10.0 & 20.0 & 1.3 & 18.7 \\
JC & 15.0 &  9.0 & 1.7 &  7.3 
\enddata
\tablenotetext{a}{Time during which the central engine is kept on} 
\tablenotetext{b}{Total energy injected at the base of the twin jets} 
\tablenotetext{c}{Estimated total ``wasted'' energy (see \S~\ref{breako})} 
\tablenotetext{d}{Estimated total energy in the twin highly relativistic
  jets at the end of the calculation (see \S~\ref{breako})}
\end{deluxetable}

\clearpage

\begin{deluxetable}{ccccc}
\tablecaption{Parameters of Models for Jet Propagation in the Stellar
  Wind \label{tab:w}} 
\tablewidth{0pt} \tablehead{ \colhead{} &
\colhead{$\dot{E}$\tablenotemark{a}} &
\colhead{$\theta_{0}$\tablenotemark{b}} & \colhead{} & \colhead{} \\
\colhead{Model} & \colhead{($10^{50}\,\ergs$)} & \colhead{(degree)} &
\colhead{$\Gamma_{0}$\tablenotemark{c}} &
\colhead{$f_{0}$\tablenotemark{d}} } 
\startdata 
W1 & 8 & 3 & 10 & 0.06 \\ 
W2 & 8 & 3 & 50 & 0.33 
\enddata
\tablenotetext{a}{Energy deposition rate} 
\tablenotetext{b}{Initial half opening angle} 
\tablenotetext{c}{Initial Lorentz factor}
\tablenotetext{d}{Initial ratio of kinetic energy to total
  energy (excluding the rest mass)} 
\end{deluxetable}

\clearpage

\clearpage

\clearpage

\clearpage

\clearpage

\clearpage

\clearpage

\clearpage

\clearpage

\clearpage

\clearpage

\clearpage

\clearpage

\clearpage

\clearpage

\clearpage

\end{document}